\documentclass[conference]{IEEEtran}
\IEEEoverridecommandlockouts
\usepackage{cite}
\usepackage{amsmath,amssymb,amsfonts}
\usepackage{ulem}
\usepackage{algorithmic}
\usepackage{graphicx}
\usepackage{textcomp}
\usepackage{xcolor}
\def\BibTeX{{\rm B\kern-.05em{\sc i\kern-.025em b}\kern-.08em
    T\kern-.1667em\lower.7ex\hbox{E}\kern-.125emX}}

\usepackage{caption,subcaption}
\usepackage{array}
\usepackage{hhline}
\usepackage{multirow}

\usepackage{hyperref}
\usepackage{graphicx}
\usepackage{placeins}

\begin{document}

\title{Know Your Account$:$ Double Graph Inference-based Account De-anonymization on Ethereum\\
}



\author{Shuyi~Miao$^{1,2,3}$ ,
        Wangjie~Qiu$^{1,2,3*}$,
        Hongwei~Zheng$^{2,4*}$,
        Qinnan~Zhang$^{1,2,3}$,
        Xiaofan~Tu$^{1,2,3}$,
        Xunan~Liu$^{1,2,3}$,\\
        Yang~Liu$^{1,2,3}$,
        Jin~Dong$^{2,4}$
        and~Zhiming~Zheng$^{1,2,3}$\\
$^1$the Institute of Artificial Intelligence, Beihang University, Beijing 100191, China\\ 
$^2$Beijing Advanced Innovation Center for Future Blockchain and Privacy Computing, \\Beihang University, Beijing 100191, China \\
$^3$Zhongguancun Laboratory, Beijing, China \\
$^4$Beijing Academy of Blockchain and Edge Computing (BABEC), Beijing, 100190, China \\
\texttt{\{shuyimiao, wangjieqiu, zhangqn, xiaofantu, xunanliu, ly9923\}@buaa.edu.cn},\\
\texttt{{\{hwzheng, zzheng\}@pku.edu.cn},
\texttt{dongjin@baec.org.cn}
}}

\maketitle

\begin{abstract}
The scaled Web 3.0 digital economy, represented by decentralized finance (DeFi), has sparked increasing interest in the past few years, which usually relies on blockchain for token transfer and diverse transaction logic. 
However, illegal behaviors, such as financial fraud, hacker attacks, and money laundering, are rampant in the blockchain ecosystem and seriously threaten its integrity and security. In this paper, we propose a novel double graph-based Ethereum account de-anonymization inference method, dubbed DBG4ETH, which aims to capture the behavioral patterns of accounts comprehensively and has more robust analytical and judgment capabilities for current complex and continuously generated transaction behaviors. 
Specifically, we first construct a global static graph to build complex interactions between the various account nodes for all transaction data.
Then, we also construct a local dynamic graph to learn about the gradual evolution of transactions over different periods. Different graphs focus on information from different perspectives, and features of global and local, static and dynamic transaction graphs are available through DBG4ETH. 
In addition, we propose an adaptive confidence calibration method to predict the results by feeding the calibrated weighted prediction values into the classifier. 
Experimental results show that DBG4ETH achieves state-of-the-art results in the account identification task, improving the F1-score by at least 3.75\% and up to 40.52\% compared to processing each graph type individually and outperforming similar account identity inference methods by 5.23\% to 12.91\%. 
\end{abstract}


\section{INTRODUCTION}
Blockchain is a decentralized, distributed public ledger that has achieved tremendous success in various industries such as finance, energy, and agriculture~\cite{27}, where well-known killer-level applications are digital cryptocurrencies~\cite{18,20}. Ethereum is one of the most representative cryptocurrencies with the largest usage and the second largest by market capitalization. Unlike traditional financial transaction systems, Ethereum's addresses are designed to be anonymous without any actual meaning. Although the anonymity of blockchain can protect user privacy, the lack of real identities also provides a protective umbrella for various financial crimes~\cite {37}. With the development of the Web 3.0 ecosystem~\cite{28}, blockchain has also introduced a series of new criminal activities, bringing new security challenges and regulatory issues~\cite{19,38,39}. According to the latest report `Hack3d' from CertiK\footnote{https://www.certik.com}, in the first quarter of 2024, there were 233 on-chain security incidents in the Web3.0 domain, resulting in losses of \$502 million, an increase of 54\% compared to the same period last year. 

To crack down on financial crime,
official governments have begun to strengthen blockchain regulation
under the anonymity of  transactions, 
as well as
researchers have deeply explored the account de-anonymization inference methods~\cite{9, 10, 13, 14, 15}. These methods most aim to mine attributes and behavioral patterns using publicly available transaction information to determine the identities behind anonymous accounts, enhancing the comprehensive understanding of the behavior of participants. Account identity inference methods for Ethereum accounts typically use graph analytics~\cite{40,9,10,14,15,16}, which models large amounts of data as graphs and treats account identification as a classification task from a graph perspective. Although graph analysis methods can enhance the learning of transaction information through deep learning, some \textbf{challenges} still need to be addressed urgently.

\begin{figure}[h]
  \centering
  \includegraphics[width=\linewidth]{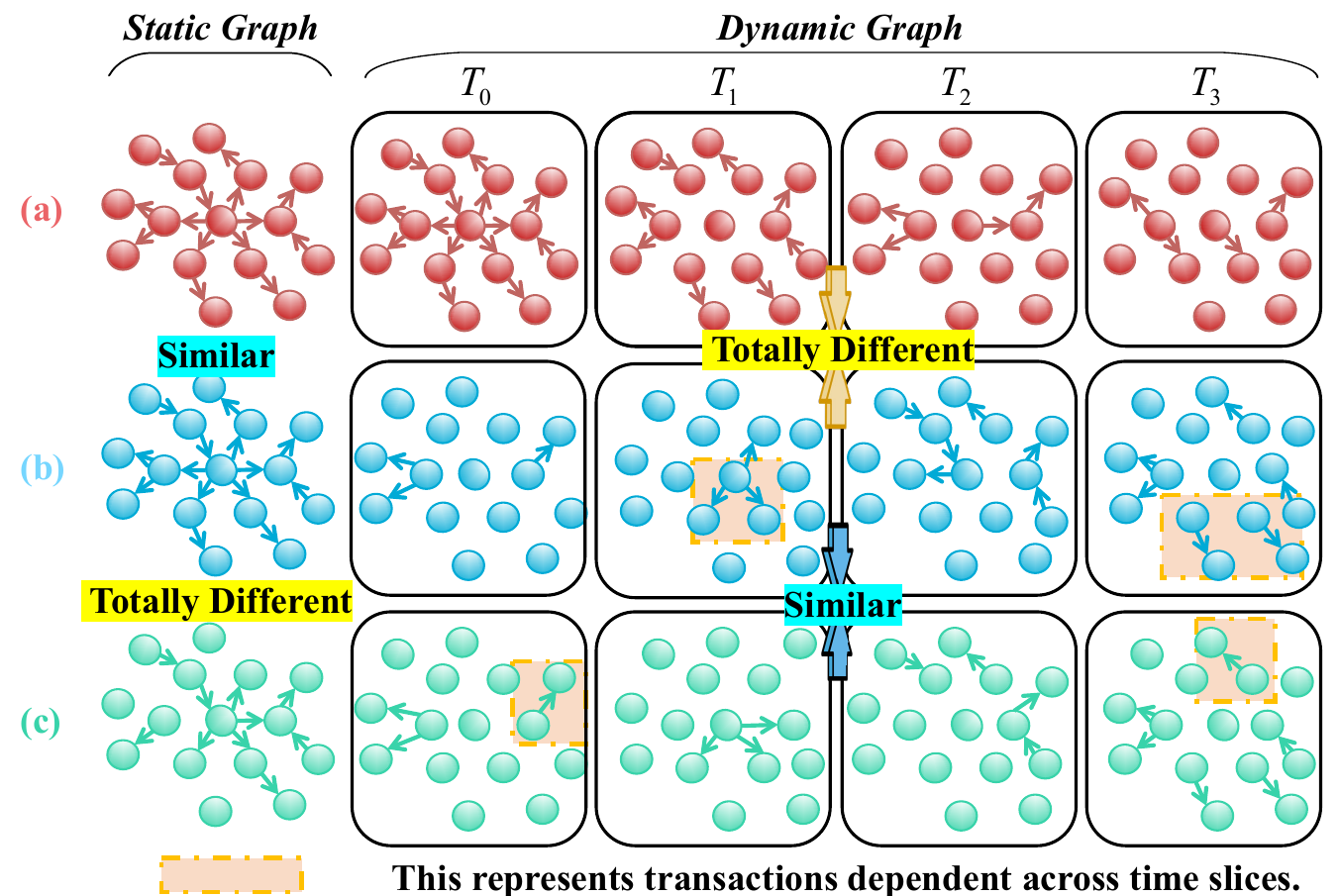}
  \caption{Details about the challenge (i): (a) and (b) have similar static graphs, but completely different dynamic graphs. (b) and (c) have similar dynamic graphs, but different static graphs. }
  \label{figure0}
\end{figure}

(i) Lack of a comprehensive graph learning model. Existing graph analysis methods discard transaction sequential information during graph construction to facilitate graph computation\cite{17}. Figure \ref{figure0} (a) and (b) show that similar static graphs may correspond to completely different evolving dynamic graphs. Analyzing account behavior only through constructing a static transaction graph will lead to a lack of detailed evolution information about the accounts. Furthermore, since transaction sampling drops out unimportant transaction information in a long perspective, even the same dynamic graph may correspond to a completely different static graph, as shown in Figure \ref{figure0} (b) and (c).

(ii) Ignoring the reliability of prediction results. Existing methods overlook the actual probability of events occurring, leading to models that may be overly confident or cautious. In particular, some scenarios may involve large-scale account de-anonymization inference and surveillance with limited resources, so higher confidence levels in predictions become exceedingly important.

(iii) The scarcity of labeled account information. The current Ethereum account labels largely depend on public platforms, such as labelcould\footnote{https://etherscan.io/labelcloud}, that provide labels according to their internal validation processes. Due to the high costs, the labor-intensive nature of the work, and the lack of a unified standard for labeling, there is a severe shortage of usable label information. This scarcity amplifies the vulnerability of tasks to underfitting or overfitting.

To address the above challenges, we propose a novel Ethereum account identity de-anonymization inference method: \uline{\textbf{D}}ou\uline{\textbf{b}}le \uline{\textbf{G}}raph inference-based account de-anonymization on \uline{\textbf{Eth}}ereum (DBG4ETH). Since directly inputting large-scale transaction graphs into graph neural networks (GNN)~\cite{6,7,8} training is not feasible, we generate account-centered subgraphs from the obtained account-centric transaction dataset, further transforming the account identification task into a subgraph-level classification task. The following describes our approach in three stages and elaborates on how we address these challenges.

\textbf{Stage 1}: Transaction data are constructed into graphs. We first select the labeled nodes as the central nodes based on the transaction data. Then, the central node is used as the center to sample the top-${K}$ nodes as neighbors according to the average transaction amount. Additionally, we filter out the remaining attributes from the transactions and generate 15-dimensional deep account features for nodes in the subgraph. These deep account features are classified into four types, including sender account features, receiver account features, transaction fee features, and contract features, which are described in detail in Section \ref{sec:section3.2}. 

\textbf{Stage 2}: Combining static-dynamic and long-short graph learning. To address the challenge (i), we propose a double graph identity inference method that combines static-dynamic and long-short terms. To be specific, we split the obtained subgraph by transaction time into multiple discrete-time dynamic graphs at fixed intervals to construct a \textbf{L}ocal \textbf{D}ynamics \textbf{G}raph (LDG), which learns the continuously evolutionary transaction information of accounts over an extended period. Meanwhile, we construct a \textbf{G}lobal \textbf{S}tatic \textbf{G}raph (GSG) to capture the long-term behavioral features of accounts, rather than being limited to the changes in features between adjacent periods captured by the LDGs. Subsequently, we obtain two prediction values from the global static account encoding module and the local dynamic account encoding module.

\textbf{Stage 3}: Adaptive confidence calibration. To solve the challenge (ii), we put two predicted values into a joint calibration module, which outputs the calibrated prediction probability values. Finally, two probability values are fed into the classifier to produce the predicted account type. To address the challenge (iii), we aim to achieve effective account classification through our elaborate designed deep features and integrated learning paradigm, even when trained on a limited amount of labeled datasets.

In summary, our main \textbf{contributions} are as follows:
\begin{itemize}
\item \textbf{DBG4ETH}: We propose a well-designed Ethereum account de-anonymization inference scheme capable of perceiving both the local dynamic changes in transactions and the long-term behavior of accounts from a more strategic perspective.
\item \textbf{Confidence}: Our method includes a confidence calibration module, which employs an adaptive confidence weight assignment method to make the model's predicted probabilities more accurately reflect the actual situation.
\item \textbf{Accurate and efficient}: The results of de-anonymization experiments on six types of Ethereum accounts demonstrate that our model can achieve state-of-the-art performance compared to multiple baselines. Moreover, our approach yields optimal results even when the label is limited. The source code is available at: \href{https://github.com/msy0513/DBG4ETH-main}{https://github.com/msy0513/DBG4ETH-main}.
\end{itemize}  

\section{BACKGROUND AND RELATED WORK}

\subsection{Ethereum Account Model}
The Ethereum account model is fundamental to the operation of the Ethereum blockchain, providing a framework for transactions and smart contract execution. 

\textbf{Accounts.} There are two primary types of accounts: externally owned accounts (EOAs), controlled by private keys, and contract accounts, which are essentially self-executing programs. EOAs enable users to send transactions, like transferring Ether or initiating smart contract functions. Contract accounts, deployed by EOAs, contain code that runs automatically upon receiving transactions, enabling complex logic and state changes on the blockchain.

\textbf{Transactions.} Transactions involve interactions between these accounts, with simple Ether transfers between EOAs and more sophisticated engagements when invoking contract account functionalities. Each account has a unique address, a balance in Ether, and a nonce for ensuring transaction ordering. This model supports a decentralized ecosystem where applications can run autonomously, enhancing security and efficiency in a trustless environment.

\subsection{De-anonymization Inference}
Ethereum account de-anonymization refers to the process of correlating blockchain transaction data with real-world entities or behaviors to uncover the identities or purposes behind ostensibly anonymous or semi-anonymous account addresses. On public blockchain networks such as Ethereum, while account addresses do not directly disclose ownership information, in-depth analysis of transaction histories, fund flow patterns, and interactions with smart contracts can reveal insights into the potential users or functions associated with these accounts. This process is of paramount importance for combating illicit activities, enhancing regulatory compliance monitoring, and advancing the analytical capabilities of the blockchain. The de-anonymization task mainly consists of several tasks such as account clustering, account identification, money laundering tracking, etc. In this paper, we focus on the Ethereum account identification task that further infers account use by tracking a user's historical transactions.

Graph representation learning methods are popular methods for dealing with Ethereum account identity inference tasks. Specifically, graph representation learning methods can better capture the identity characteristics of anonymous accounts by constructing massive transaction data as a graph and studying the topological relationships among accounts. Related studies are presented in the following Section ~\ref{sec:section2}. Besides using graph learning to implement Ethereum account identity inference, some methods apply language models. BERT4ETH~\cite{17} employs a universal pre-trained Transformer~\cite{21} encoder to capture dynamic temporal information in transactions, further detecting fraud activities.

\subsection{Graph Representation Learning}
\label{sec:section2}
The account identity inference task usually uses graph representation learning methods which can be mainly divided into graph embedding methods and GNN-based Methods.
\subsubsection{Graph Embedding Methods}
Traditional graph embedding methods typically map nodes in the graph to low-dimensional vectors and then use machine learning models for classification and prediction. Random walks are a common graph traversal technique used to explore sequences of nodes randomly within a graph. For example, DeepWalk~\cite{1} uses random walks to sample sequences of nodes, employing the Word2Vec~\cite{2} method to learn the low-dimensional vector representations of nodes. Node2Vec~\cite{3} evolves from DeepWalk by introducing two parameters ${p}$ and ${q}$, to implement biased random walks, thereby generating training sequences for nodes. In specific blockchain applications, Yuan et al.~\cite{4} proposed using random walk methods on graphs to capture behavioral patterns of accounts in transaction graphs. Trans2Vec~\cite{5} enhances random walks on graphs by incorporating transaction amounts and timestamps to extract compelling features for completing phishing detection tasks.

\subsubsection{GNN-based Methods}
Graph embedding methods can directly compress data, making vector calculations simpler and faster than operating directly on graphs. However, the method is unable to optimize the quality of embeddings based on feedback from classifiers. GNNs offer a solution that can process graph data and produce predictive results end-to-end. 

Traditional GNN-based methods, such as graph convolutional network (GCN)~\cite{6}, adapt the convolution operation from image processing to graph-structured data for the first time, allowing nodes to aggregate and update information from their neighbors to learn node representations. Graph attention network (GAT)~\cite{7} introduces an attention mechanism, allowing nodes to dynamically weigh different neighbors based on their importance when aggregating neighbor information, enhancing the model's flexibility and expressiveness. Graph isomorphism network (GIN)~\cite{8} design a new aggregation mechanism to capture topological structures on graphs, effectively improving the learning capabilities of graph structures.

\subsection{GNN-based De-anonymization Methods}
In GNN-based blockchain account identity inference tasks, $\mathrm{I}^{2}\mathrm{BGNN}$~\cite{9} learns to map transaction subgraph patterns to account identities, achieving account de-anonymization with subgraphs as inputs. Ethident~\cite{10} is an end-to-end graph neural network framework for account de-anonymization, which designs 49 graph enhancement methods for different categories of accounts. RiskProp~\cite{13} proposes an account risk rating method based on directed bipartite graphs to quantify the distribution of risks in transaction networks. KYC-GCN~\cite{14} proposes using an improved GCN architecture to handle multiple aggregators and data based on importance sampling, to achieve account category inference. Unlike the above methods that construct static graphs for identity inference, some works use dynamic graphs for phishing detection tasks. TEGDetector~\cite{15} transforms transaction sequences into multiple time slices and learns transaction behaviors across different time intervals according to time coefficients. Additionally, some methods consider the heterogeneity of graph nodes to solve the Ethereum account classification problem. BPA-GNN~\cite{16} considers both isomorphic and heterogeneous features during neighbor aggregation to obtain the final prediction.

\section{DOUBLE GRAPH CONSTRUCTION}
\subsection{Problem Definition}
We define two account interaction graphs. GSG is represented by 
${G_{g}=\{V, E, X, R, Y\}}$, where ${V=\{v_{1}, v_{2}, ..., v_{n}\}}$ denotes the set of account nodes, ${E\subseteq {\{(v_{i},v_{j}})|v_{i},v_{j}\in{V}\}}$ is the interaction edges between the nodes, ${\mathbf{X}\in{\mathbb{R}^{n\times d_{1}}}}$ is the node feature matrix, and ${\mathbf{R}\in{\mathbb{R}^{m\times d_{2}}}}$ is the edge feature matrix. ${n}$ denotes the total number of nodes, and ${m}$ denotes the total number of edges. $Y=\{(v_{i},y_{i})\mid v_{i}\in V\}$ denotes the label set of the corresponding account nodes.

LDG can be denoted as ${G_{l}=\{V, E, X, R, T, Y\}}$, where ${T}$ denotes the transaction evolution time and the remaining notation is the same as the GSG definition. Assume that ${E_{i}=\{e_{0}, ..., e_{k}\}}$ is the set of ${k}$ neighbouring transactions extracted centred on the target address ${v_{i}}$, and the transaction evolution time of ${e_{j}\in{E_{i}}}$ can be defined as follows:
\begin{equation}
  \ T_{e_{j}}(V_{i},E_{i})=\frac{t_{j}-t_{min}}{t_{max}-t_{min}}
\end{equation}
where ${V_i}$ is the set of all nodes in subgraph centred on ${v_i}$, ${t_{j}}$ is the timestamp of transaction edge ${e_{j}}$, ${t_{min}}$ and ${t_{max}}$ denote the minimum and maximum values of the transaction evolution time in ${E_{i}}$, respectively. We use the transaction evolution time to partition the LDG into ${T}$ time slices, where ${G_{i,k}=\{V_i, E_{i,k}\}}$ denotes the transaction graph under the ${k}$-th time slice for ${v_{i}}$ centred transactions. Therefore, the blockchain network can be modeled as a dynamic graph classification dataset, including ${N}$ graphs. 

\subsection{Ethereum Data Processing}
\subsubsection{Transaction Data Filtering} We first obtain Ethereum transaction data and labeled account information. Secondly, we delete all unsubmitted transactions. Then, we sample the top-${K}$ important neighbors for the target account ${v_{i}}$ based on the average transaction value. For each of neighbors, we also sampled their top-${K}$ important neighbors. The entire sampling process can be described as follows:
\begin{equation}
    V_k=\bigcup_{v\in V_{k-1}}topK\left(\mathcal{N}_{\nu},K,\mathbf{R}\left[v,\mathcal{N}_{\nu},i\right]\right),i\in\{0,1,2\}
\end{equation}
where \( V_k \) is the set of nodes obtained through ${k}$-hop sampling (${k=0}$ indicates ${v_i}$), ${\mathcal{N}_{\nu}}$ is the set of nodes corresponding to the edges within 1-hop that are sampled, $K$ is the number of neighbors selected for each edge, ${\mathbf{R}\left[v,\mathcal{N}_{\nu},i\right]}$ guides which edge to sample based on the average transaction value, and ${i}$ guides the specific edge to select. ${V_i=\cup_{k=0}^hV_k}$ represents the set of nodes constructed by sampling ${h}$-hop neighbors from the target account ${v_i}$. If there are duplicate average transaction amounts in the sample, we will choose the transaction information based on the total transaction value.

\subsubsection{Node Feature Construction}
\label{sec:section3.2}
To leverage the obtained transaction data, we add 15-dimensional deep features to the account nodes for forming a broader statistical feature space, which can be classified into the following four categories, as shown in Table \ref{t2}.
\begin{table}[h]
  \fontsize{7.5}{11.5}\selectfont 
  \setlength{\tabcolsep}{0.5pt} 
\centering
\caption{The extraction of 15-dimensional deep features on transactional data.}
\label{t2}
\begin{tabular}{|@{}c|c|c|@{}|}
\hline
\textbf{Deep feature types}& \textbf{Abbreviations} & \textbf{Definition} \\ \hline
\multirow{4}{*}{\begin{minipage}[c]{2cm}
                  \centering
                  \textbf{Sender account}\\
                  \textbf{features}
                \end{minipage}} 
 & NTS & Number of Transactions Sent \\ 
& STV & Send Total Value \\
& SAV & Send Average Value \\
& min/max\_STI & Minimum /Maximum Send Time Interval \\ \hline
\multirow{4}{*}{\begin{minipage}[c]{2cm}
                  \centering
                  \textbf{Receiver account}\\
                  \textbf{features}
                \end{minipage}} 
 & NTR & Number of Transactions Received \\ 
& RTV & Receive Total Value \\
& RAV & Receive Average Value \\
& min/max\_RTI & Minimum /Maximum Receive Time Interval \\ \hline
\multirow{2}{*}{\begin{minipage}[c]{2cm}
                  \centering
                  \textbf{Transaction}\\
                  \textbf{fee features}
                \end{minipage}} & SETF/RETF & Send/Receive Ether Transaction Fee \\ 
& SAETF/RAETF & Send/Receive Average Ether Transaction Fee \\ \hline
\textbf{Contract feature} & NC & Number of Contract calls \\ 
\hline
\end{tabular}
\end{table}

\textbf{Sender account features} include the number of transactions sent from the account (NTS), the total value of transactions sent from the current account (STV), the average value of transactions sent from the account (SAV) and the interval between two consecutive transactions sent from the account (STI). We use max{\_}STI and min\_STI to denote the maximum and minimum values of the send transaction interval. For example, ${T_{i,k}}$ denotes the timestamp of the ${k}$-th transaction sent by the ${i}$-th account, max{\_}STI and min{\_}STI can be expressed as:
\begin{equation}max\_STI_i=\max_{k}(|T_{i,k}-T_{i,k+1}|)\end{equation}
\begin{equation}min\_STI_i=\min_{k}(|T_{i,k}-T_{i,k+1}|)\end{equation}

\textbf{Receiver account features} include the number of transactions received (NTR), receive total value (RTV), receive average value (RAV), max receive time interval (max\_RTI), and min receive time interval (min\_RTI), which are calculated similarly to the features of sender transactions account features.

\textbf{Transaction fee features} include the total Ether transaction fee for each account (TETF) and the average Ether transaction fee (AETF). For the STETF of $i$-th account as the sender can be expressed as follows:
\begin{equation}
    STETF_i=\sum^{NTS_i}_{j=1}(gasPrice_{i,j}\times gasUsed_{i,j})\times10^{-18}
\end{equation}
where ${10^{-18}}$ denotes the uniform conversion of the smallest unit of Ether, Wei, into ETH.

An account in Ethereum is an entity that owns Ether, which can be divided into two categories: EOA and contract account (CA). With the transaction data, we can determine whether the sender and receiver of this transaction are an EOA or a CA. \textbf{Contract feature} is the total number of times all contracts (NC) are called in transactions involving each account.

\subsubsection{Interaction Merging and Edge Feature Construction} 
Transactions from the source account ${v_{i}}$ to the target account ${v_{j}}$ are merged into one edge, which is characterized by the total amount $w$ and the total number of times of these ${t}$ transactions. For the GSG, the features of edge from  ${v_{i}}$ to ${v_{j}}$ can be expressed as \( r_{ij} = [w, t] \). For LDG, we use the total amount of transactions as the features of the edges. The feature of edge from  ${v_{i}}$ to ${v_{j}}$ in the ${k}$-th time slice can be denoted as \( r_{ij}^{k} = [w^k] \).

\section{METHODOLOGY}
\begin{figure*}[htbp]
  \centering
  \includegraphics[width=\linewidth]{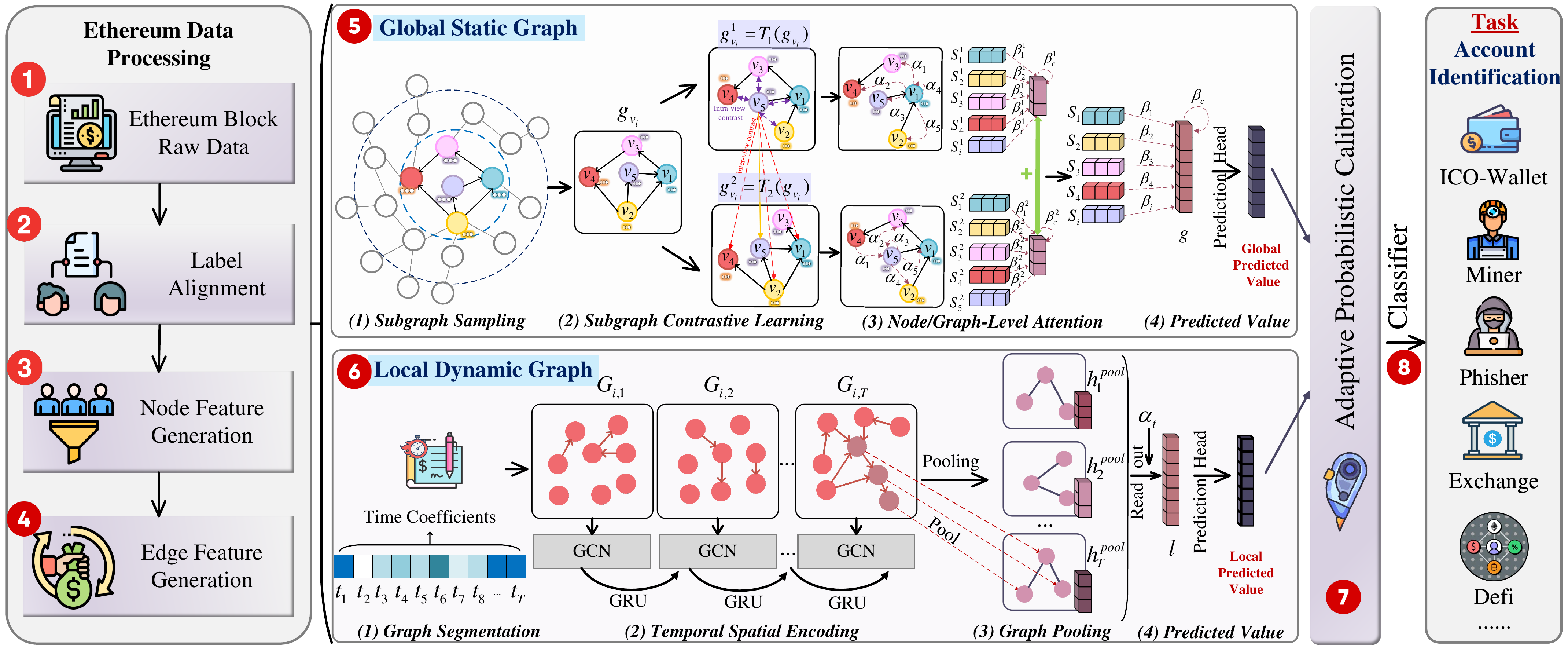}
  \caption{The overview of our DBG4ETH.}
  \label{figure1}
\end{figure*}

We describe the de-anonymized graph inference network DBG4ETH in detail as shown in Figure \ref{figure1}. For the target account ${v_{i}}$, the inputs of DBG4ETH are processed node features and edge features, and the outputs are predicted account identity labels. DBG4ETH consists of the following four components: (1) global static account transaction encoding module, (2) local dynamic account transaction encoding module, (3) joint prediction and calibration module, and (4) account classification module. Next, we describe each component in detail.

\subsection{Global Static Account Transaction Encoding Module}
To thoroughly learn the account behavior patterns on the graph, we employ GNN and a hierarchical attention network~\cite{11} for learning the graph's topological structure. Although transaction data on Ethereum is readily available and abundant, there is currently a scarcity of related labeled information. To mitigate the sparsity of account labels and enhance the capability of graph embedding learning, we introduce contrastive learning with adaptive augmentation~\cite{12} as a regularization tool to jointly train the learning of graph topological structures. In the following, we will describe the specific encoder for GSG.

\subsubsection{Node feature alignment}
To facilitate processing by GNN and better preserve the local structure of the graph, we fuse the features of each neighbor node \( v_j \) in the subgraph centered at node \( v_i \), with the corresponding connecting edge information \( r_{ij} \) to represent the neighbor node's features as \( [x_j || r_{ij}] \). Therefore, we use linear transformations and activation functions to eliminate the dimensionality differences between nodes to align features of different dimensions. We employ a linear layer and ${LeakyReLU}$ \cite{46} to update and align the feature dimensions of neighbor nodes:
\begin{equation}
\tilde{\mathbf{x}}_j=LeakyRelu\left(\Theta_x\cdot[\mathrm{x}_j\parallel\mathrm{r}_{ij}]\right)
\end{equation}

\subsubsection{Hierarchical attention network}
Our goal with the hierarchical graph attention network~\cite{11} is to find the most relevant parts of the neighborhood to generate a representation for center node \( v_i \). Hierarchical graph attention network has two main components: node-level attention mechanism and graph-level graph attention mechanism. Therefore, for the node-level graph attention mechanism, as shown in Figure \ref{figure1} part 5(3), for each node with different colors in the subgraph, we use the graph attention mechanism (GAT)~\cite{7} to select more relevant neighboring nodes and give them higher weights and finally update the representation of each node with different colors by combining the neighbor information. Specifically, for node ${v_i}$ in the subgraph, the node-level graph attention mechanism is used to learn the attention scores of the hidden features of the neighbors ${v_j}$ in layer ${l}$ as follows:
\begin{equation}
s_{ij}^l=LeakyRelu\left(\Theta_n^l\cdot[{H}_i^l\parallel{H}_j^l]\right)
\end{equation}
where ${\Theta_n^l}$ is the linear transformation parameter. Next, the neighbor importance scores are normalized for each layer:
\begin{equation}
    \alpha_{ij}^{l}=Softmax\left(s_{ij}^{l}\right)=\frac{\exp\left(s_{ij}^{l}\right)}{\sum_{x\in{V}_i^l}\exp\left(s_{ix}^{l}\right)}
\end{equation}
where ${V_i^l}$ denotes the ${l}$-hop neighbors of ${v_i}$. Finally, we use the normalized attention scores for each layer to update the account features by aggregating the neighbor information:
\begin{equation}
H_i^{l+1}=Elu\left(\alpha_{ii}^l\cdot\Theta_\alpha^l\cdot{H}_i^l+\sum_{j\in{V}_i^l}\alpha_{ij}^l\cdot\Theta_\alpha^l\cdot{H}_j^l\right)
\end{equation}
where ${\Theta_{\alpha}^{i}}$ is a linear layer. Starting from the ${h}$-th layer neighbors of the account node ${v_i}$ and updating inward consecutively, the final output ${H_{i}^{h}}$ represents the embedded features of the target node, which includes information from all neighbors within ${h}$-hops. 

After obtaining the account node representations with different colors in the subgraph, as shown in Figure \ref{figure1} part 5(3), it is necessary to aggregate the nodes in the graph to generate a graph representation of the subgraph representations. Unlike previous methods that directly use mean, sum, or maximum pooling to aggregate node features, we first obtain the initial subgraph representation using global maximum pooling:
\begin{equation}
    c=MaxPooling\left(H^h_{i}\right)
\end{equation}
Subsequently, we utilize GAT to learn the attention score ${s_j}$ for any node ${v_j}$ in the subgraph towards the overall graph representation:
\begin{equation}
s_j=LeakyRelu\left(\Theta_s\cdot[\mathrm{c}\parallel{H}_j^h]\right)
\end{equation}
where ${\Theta_s}$ is the linear layer. The graph-level graph attention scores are normalized:

\begin{equation}
    \beta_j=Softmax\left(s_j\right)=\frac{\exp(s_j)}{\sum_{x\in V_i\cup\{c\}}\exp(s_x)}
\end{equation}
where ${s_c}$ is the self-attention score of ${c}$. Finally, subgraph embedding can be represented as:
\begin{equation}
 g=Elu\left(\beta_c\cdot\Theta_g\cdot{c}+\sum_{j\in V_i}\beta_j\cdot\Theta_g\cdot {H}_j^h\right)
\end{equation}
where ${\Theta_{g}}$ is a linear layer.

\subsubsection{Contrastive Learning with Adaptive Augmentation}
We use graph contrast learning with adaptive augmentation ~\cite{12} to randomly generate two viewpoint graphs ${g_{\nu_i}^1}$ and ${g_{\nu_i}^2}$ using a graph augmentation function on the input subgraph ${g_{\nu_i}}$, and then use a contrasting objective to force the representations of the same graphs in the two different viewpoints to be as similar as possible, and as far away from each other as possible concerning the representations of the different graphs.

Inspired by~\cite{12}, we draw on its adaptive graph augmentation function, which is an augmentation scheme that can add disturbances to unimportant edges and features, and important topology and node features can also be maintained after augmentation by the graph. Specific graph augmentation methods include topology-level augmentation and node-attribute-level augmentation. (1) Topology-level augmentation: edge centrality is computed based on node centrality, and edges with low edge centrality are eliminated. The node centrality function is determined by three measures: degree centrality~\cite{41}, eigenvector centrality~\cite{41}, and PageRank centrality~\cite{41}. (2) Node-attribute-level augmentation: node attribute augmentation by randomly masking a fraction of dimensions with zeros in node features. We note the labels obtained for the generated graph as the labels of the original graph as well. Subgraph contrast learning on the GSG maximizes the consistency of the two augmented subgraphs in the contrast space by minimizing the self-supervised loss. By adding contrast learning, the original and augmented graphs are all used to train the model, which greatly improves the model's robustness in different scenarios.

\subsection{Local Dynamic Account Transaction Encoding Module}
To better encode the intra-graph topological information and inter-graph temporal dependencies of LDG, the encoding module uses GNN to learn the complex interactions within each graph and gated recurrent unit (GRU)~\cite{25} to learn the evolutionary information between different time slices.

Specifically, we first feed the LDG into a GCN to learn the node representations ${U_t}$ for the graph at time slice ${t}$ as topological features:
\begin{equation}
    U_t=\mathrm{GCN}(h_{t-1},A_t)
\end{equation}
where ${A_t}$ is the adjacency matrix of LDG at time slice ${t}$, and $h_{t-1}$ represents the evolutionary features from the previous time slice ${t-1}$, with the initial ${h_0}$ being the input node features. Then, the evolutionary features are updated by GRU. Specifically, based on the current topological features $U_{t}$ and the previous evolutionary features $h_{t-1}$, the update gate ${u_t}$ and reset gate ${r_t}$ are computed as follows:
\begin{equation}
    u_{t}=\sigma(U_tW_u+h_{t-1}V_u)
\end{equation}
\begin{equation}
    r_{t}=\sigma(U_tW_r+h_{t-1}V_r)
\end{equation}
where ${W_u}$, ${V_u}$ and ${W_r}$, ${V_r}$ are the weight matrices of the update gate and reset gate. The update gate determines how much of the evolutionary information $h_{t-1}$ from the previous time slice is passed to the next time slice to be learnt, and the reset gate determines how much of the evolutionary information $h_{t-1}$ from the previous time slice is forgotten. The reset gate is used to compute the candidate evolutionary features for time slice ${t}$:
\begin{equation}
    \tilde{h}_t=\tanh(WU_t+(r_t\odot h_{t-1}V))
\end{equation}
where ${W}$ and ${V}$ denote the weight matrices of the reset gate, and ${\odot}$ denotes the hadamard product. Finally, the local dynamic account encoder updates the evolutionary features under ${t-1}$ time slice based on the evolutionary features $h_{t-1}$ and the candidate evolutionary features ${\tilde{h}_t}$:
\begin{equation}
    h_t=(1-u_t)\odot h_{t-1}+u_t\odot\tilde{h}_t
\end{equation}

After obtaining the structural and temporal features within the subgraphs, we use the graph pooling method Diffpool~\cite{26} to learn the clustering assignment matrix for multiple dynamic graphs. Specifically, for the ${t}$-th time slice, the clustering assignment matrix ${M_t}$ for that current time slice ${t}$ is first accelerated:
\begin{equation}
    M_t=Softmax\big(GNN(A_t,h_t)\big)
\end{equation}

The generated matrix ${M_t}$ represents the allocation of all the ${N}$ nodes initial to the ${t}$-th time slice to the new ${N\times r}$ nodes according to the allocation rate ${r}$. The evolutionary feature ${h_t^{\mathrm{pool}}}$ and the adjacency matrix ${A_t^{\mathrm{pool}}}$ can be expressed as:
\begin{equation}
    h_t^{\mathrm{pool}}=M_t^Th_t\in\mathbb{R}^{N'\times d}
\end{equation}
\begin{equation}
    A_t^{\mathrm{pool}}=M_t^TA_tM_t\in\mathbb{R}^{N'\times N'}
\end{equation}
where ${N'}$ is the number of node clusters after pooling. The evolutionary features ${\{\mathrm{h}_1^{\mathrm{poo}1}, \mathrm{h}_2^{\mathrm{pool}}, ...,\mathrm{h}_T^{\mathrm{pool}}\}}$ and adjacency matrices ${\{\mathrm{A}_1^{\mathrm{pool}}, \mathrm{A}_2^{\mathrm{pool}}, …, \mathrm{A}_T^{\mathrm{pool}}\}}$ can be obtained by pooling. To obtain a unique representation of the central node ${v_i}$, we use the \textit{Read-out} \cite{47} operation to add time slice weights to the evolutionary features of each time slice to make the important time slice more useful for the final node identity judgement. The target node ${v_i}$ can be represented as:

\begin{equation}
    \gamma_i=\sum_{t=1}^T\alpha_th_t^\text{pool}
\end{equation}
where ${\alpha_t}$ is the value of adaptive weights assigned to each time slice automatically learnt by the model. Finally, the predicted value of the LDG for the identity of node ${v_i}$ as follows:
\begin{equation}
    l=\mathrm{ReLu}(\mathrm{\Theta_{g}}(\gamma_i))
\end{equation}
where ${\Theta_{g}}$ is the linear layer.

\subsection{Joint Prediction and Calibration Module}
\label{sec:section3.3}
\begin{figure}[h]
  \centering
  \includegraphics[width=\linewidth]{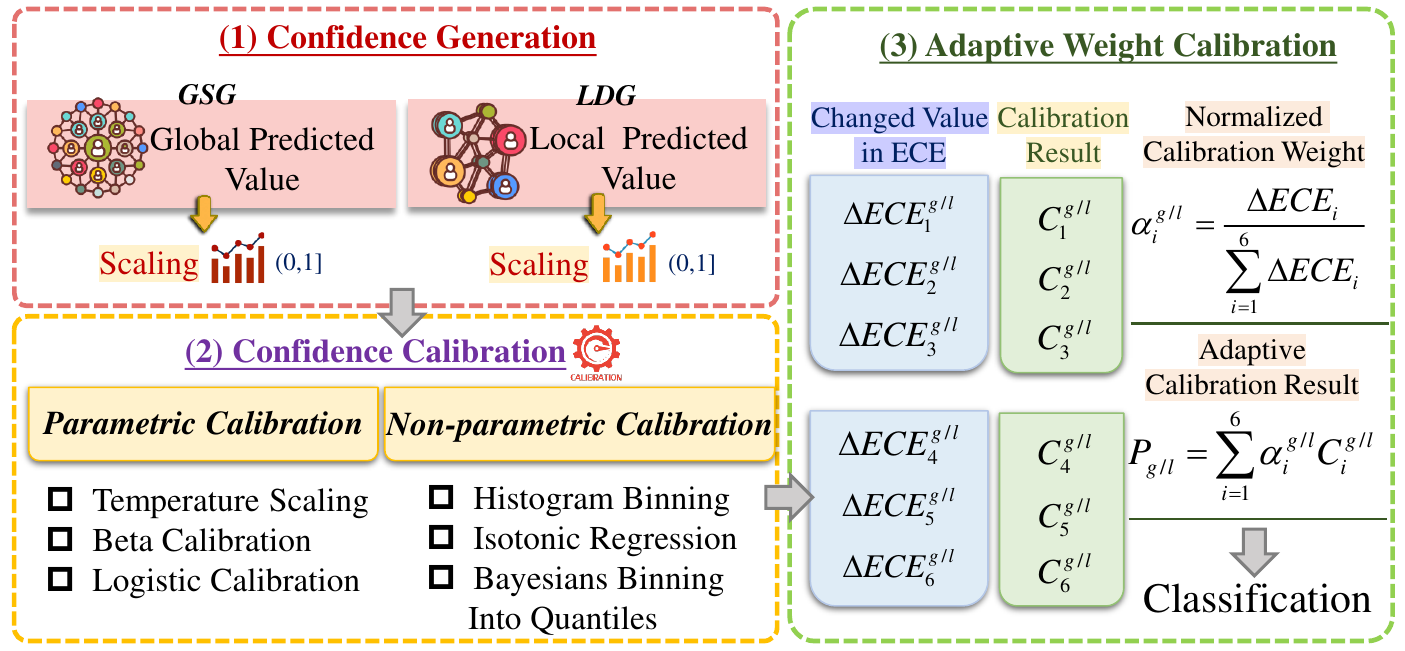}
  \caption{Adaptive calibration method for two types of prediction values: global static graph and local dynamic graph.}
  \label{figure2}
\end{figure}

After processing through global static and local dynamic graph encoders, we obtain two types of predictive values representing account predictions from different perspectives. However, since different graphs have different complex networks, too much neural network stacking can bring overconfidence problems, and the confidence of classification results may be overestimated. Meanwhile, in the actual Ethereum account classification task, inferring the account type only from the predicted probability categories is not entirely trustworthy, and it is also necessary to provide confidence corresponding to the fact. Therefore, instead of simply combining and processing the predictions obtained from the GSG and LDG, we employ post-processing calibration and an elaborate adaptive prediction value calibration method for the predictions under two parallel branches. As shown in Figure \ref{figure2}, the joint prediction and calibration module consists of the following three stages, with ${g/l}$ denoted as GSG or LDG.

\subsubsection{Confidence generation} Because two prediction results ${g}$ and ${l}$ are not in the range of 0 to 1, we first scale the predicted values according to their mean and standard deviation. Then, we obtain the predicted values that fit into the range of the two models' confidence values, respectively. 

\subsubsection{Confidence calibration} Existing calibration methods are mainly divided into two categories: parametric and non-parametric calibration methods. Parametric calibration methods can efficiently take advantage of the data and calibrate the model using limited data to guarantee classification accuracy. In contrast, non-parametric calibration methods provide a more realistic estimate of uncertainty in the presence of unknown data distributions and complete calibration without increasing model complexity and training costs. We selected three parametric and three non-parametric calibration methods to calibrate the two parallel branches, respectively. The parametric calibration methods include temperature scaling~\cite{22}, beta calibration~\cite{35} and logistic calibration~\cite{35}. The non-parametric calibration methods include histogram binning~\cite{32}, isotonic regression~\cite{33} and bayesian binning into quantiles (BBQ)~\cite{34}.

\subsubsection{Adaptive weight calibration} The evaluation metric of confidence calibration is expected calibration error (ECE)~\cite{22}, which measures the calibration performance of the model by comparing the difference between the model's predicted value and the real value. A smaller value of ECE indicates that the model's predicted value is more acceptable or credible, and a larger value indicates that the model's predicted result is less acceptable or credible.

Since parametric calibration methods improve the ability to calibrate well on limited data and non-parametric calibration methods improve the overall model classification ability, we propose a confidence calibration method that incorporates both types of multiple methods. We use this before-and-after difference to indicate the reduction in ECE, and if the value is larger, the calibration method is applied more efficiently. We calculate the ECE reduction under different calibration methods for GSG and LDG separately. For the same model, if the ECE reduction of a calibration method is larger, the calibration result obtained by that method should get a larger weight score. Thus, the calibration results can be expressed as follows:
\begin{equation}
    P_{g/l}=\sum_{i=1}^6\alpha^{g/l}_iC^{g/l}_i
\end{equation}
where ${P_{g/l}}$ represents the weighted calibration result of the GSG or LDG, ${C^{g/l}_i}$ is the calibration result of the ${i}$-th calibration method on the GSG or LDG, and ${\alpha^{g/l}_i}$ is the normalized weight obtained by the ${i}$-th calibration method on the GSG or LDG:
\begin{equation}
    \alpha_i^{g/l}=\frac{\Delta ECE_i}{\sum_{i=1}^6\Delta ECE_i}
\end{equation}
where ${\Delta ECE_i}$ is the changed value in ECE for the ${i}$-th calibration result.

\subsection{Account Classification Module}
The account classification model aims to synthesize the long-view analysis capability in the GSG and the evolutionary perception capability in the LDG to accurately classify accounts. LightGBM~\cite{23} is a gradient boosting-based decision tree method that is robust to data outliers and noise and thus performs well in the classification task. Therefore, we choose LightGBM to classify the calibration results of global static account graphs and local dynamic account graphs.

\section{EXPERIMENTS}
In this section, we validate the practical performance of the proposed method, DBG4ETH, addressing the following research questions:
\begin{itemize}
\item {RQ1: What are the selected components of DBG4ETH?}
\item {RQ2: How does our framework DBG4ETH perform compared to the current state-of-the-art Ethereum account deanonymization methods?}
\item {RQ3: How do different modules affect the performance of the proposed method?}
\item {RQ4: How robust is the model in predicting the identities of novel account types that may arise in the dynamic cryptocurrency market?}
\item {RQ5: Is DBG4ETH sensitive to hyperparameters? How do significant hyperparameters affect the performance of the model?}
\end{itemize}
\subsection{Experimental Setup}
\subsubsection{Datasets}
Following the method of data acquisition in~\cite{10}, we obtain the block data (the time interval is between “2015-08-07” to “2024-02-18”) in the block transaction column of the Ethereum On-chain Data from the Xblock website\footnote{http://xblock.pro/}, obtain the labeled account information through XLabelCloud\footnote{https://xblock.pro/\#/labelcloud} and Etherscan Lable Cloud\footnote{https://etherscan.io/labelcloud}. We match the labeled accounts with all participating accounts in the transactions and retain the top-K transactions related to the labeled accounts based on the average transaction value. The other accounts in these filtered transactions are considered the first-order neighbors for constructing subgraphs. We can obtain k-hop subgraphs centered on the known labeled accounts by iterating this process.

The dataset details are shown in Table \ref{tab:3}, where the number of positive samples indicates the number of labeled accounts and the graph denotes the number of graphs containing positive and negative examples. We select the four categories of exchange, ico-wallet, mining and phish/hack, which account for a high percentage of the total number of 2,433. In addition, to validate our model's capability to handle novel account types, we add two types: defi and bridge. Bridge accounts facilitate interoperability between different blockchain networks by enabling the transfer of assets or information across them. DeFi accounts refer to the interaction with decentralized finance applications on Ethereum.

\begin{table}[htbp]
  \fontsize{7}{11.5}\selectfont 
  \setlength{\tabcolsep}{0.8pt} 
  \caption{Dataset information for the six types of accounts.}
  \label{tab:3}
  \begin{tabular}{|c|c|c|c|c|c|c|}
    \hline
    \textbf{Dataset}&\multicolumn{6}{c|}{\textbf{Label}} \\
    \cline{2-7} 
    \textbf{information} & \textbf{Exchange} & \textbf{ICO-Wallet} & \textbf{Mining} & \textbf{Phish/Hack} &\textbf{Bridge}  &
    \textbf{DeFi}\\
    \hline
    \textbf{Num. of positive samples} & 231 & 155& 56 & 1991 &105 &105\\
    \hline
    \textbf{Graph} & 460 & 310 & 110 & 2430 &210 &210\\
    \hline
    \textbf{Ave. Num. of nodes} & 92.97 & 84.62 & 101.77 & 77.35 &119.42 &83.59\\
    \hline
    \textbf{Ave. Num. of edges} & 205.80 & 178.34 & 232.09 & 163.39 &219.01 &194.37 \\
    \hline
  \end{tabular}
\end{table}

\subsubsection{Evaluation metrics}
 We evaluate the Ethereum account identification task's performance with Precision, Recall, F${_1}$ and Accuracy.
 
\subsubsection{Baselines}
In our study, we compared a total of 14 baseline models across three categories.

Graph embedding methods:
\begin{itemize}
\item {DeepWalk~\cite{1}: It employs random walks to generate node sequences, which are then used to learn node embeddings through a Skip-Gram model.}
\item {Node2Vec~\cite{3}: It extends DeepWalk by balancing local and global exploration to learn more nuanced node embeddings.}
\end{itemize}

GNN-based methods:
\begin{itemize}
\item {GCN~\cite{6}: It uses graph convolutions to aggregate node features from the neighborhood for node representation learning.}
\item {GAT~\cite{7}: It applies attention mechanisms to weigh the importance of neighboring nodes during feature aggregation.}
\item {GIN~\cite{8}: It leverages the graph isomorphism principle to learn node embeddings that are invariant to graph permutations.}
\item {GraphSAGE~\cite{45}: It employs a sampling-and-aggregation framework to efficiently generate node embeddings from large graphs.}
\item {APPNP~\cite{36}: It approximates personalized PageRank scores to capture node relationships for representation learning.}
\item {GRIT~\cite{43}: It integrates graph inductive biases into Transformers, enabling effective node representation learning without explicit message passing.}
\end{itemize}

Deanonymization in Ethereum:
\begin{itemize}
\item {Trans2Vec~\cite{5}: It captures the temporal and relational dynamics of transactions through a combination of random walks and a deep learning framework to produce embeddings for de-anonymization tasks.}
\item {$\mathrm{I}^{2}\mathrm{BGNN}$~\cite{9}: It iteratively refines node representations using graph inductive biases without explicit message passing.}
\item {TSGN~\cite{44}: It captures temporal and subgraph dynamics to learn embeddings for evolving graph structures.}
\item {Ethident~\cite{10}: It utilizes a hierarchical graph attention mechanism to identify patterns and behaviors for Ethereum account de-anonymization.}
\item {TEGDetector~\cite{15}: It analyzes temporal event graphs to detect abnormal accounts.}
\item {BERT4ETH~\cite{17}: It is a universal pretrained Transformer encoder designed to extract illegal account representations on the Ethereum blockchain.}
\end{itemize}

\subsubsection{Parameter settings}
\label{sec:section5.1}
In the sampling process of constructing the graph, we set the maximum number of sampling hops to 2 and the maximum number of sampling transactions per hop ${K=2000}$. For GSG, adaptive contrast learning generates two graphs where we set the ratio of randomly removed node features to 0.1 and 0.0, and the corresponding ratios of randomly removed edges to 0.3 and 0.4, respectively. We set the encoder to 2-layer GAT, the hidden layer to 128, and the pooling method to maximum pooling. For LDG, we set ${T=10}$, the number of pooling to 2, and the number of address clusters after two pooling is ${N_1^{\prime}=N*0.1}$ and ${N_2^{\prime}=1}$. During training, we use Adam optimizer to find the optimal learning rate among \{0.1,0.05,0.01,0.005,0.001\}.


\begin{figure}[htbp]
  \centering
  \includegraphics[width=8cm]{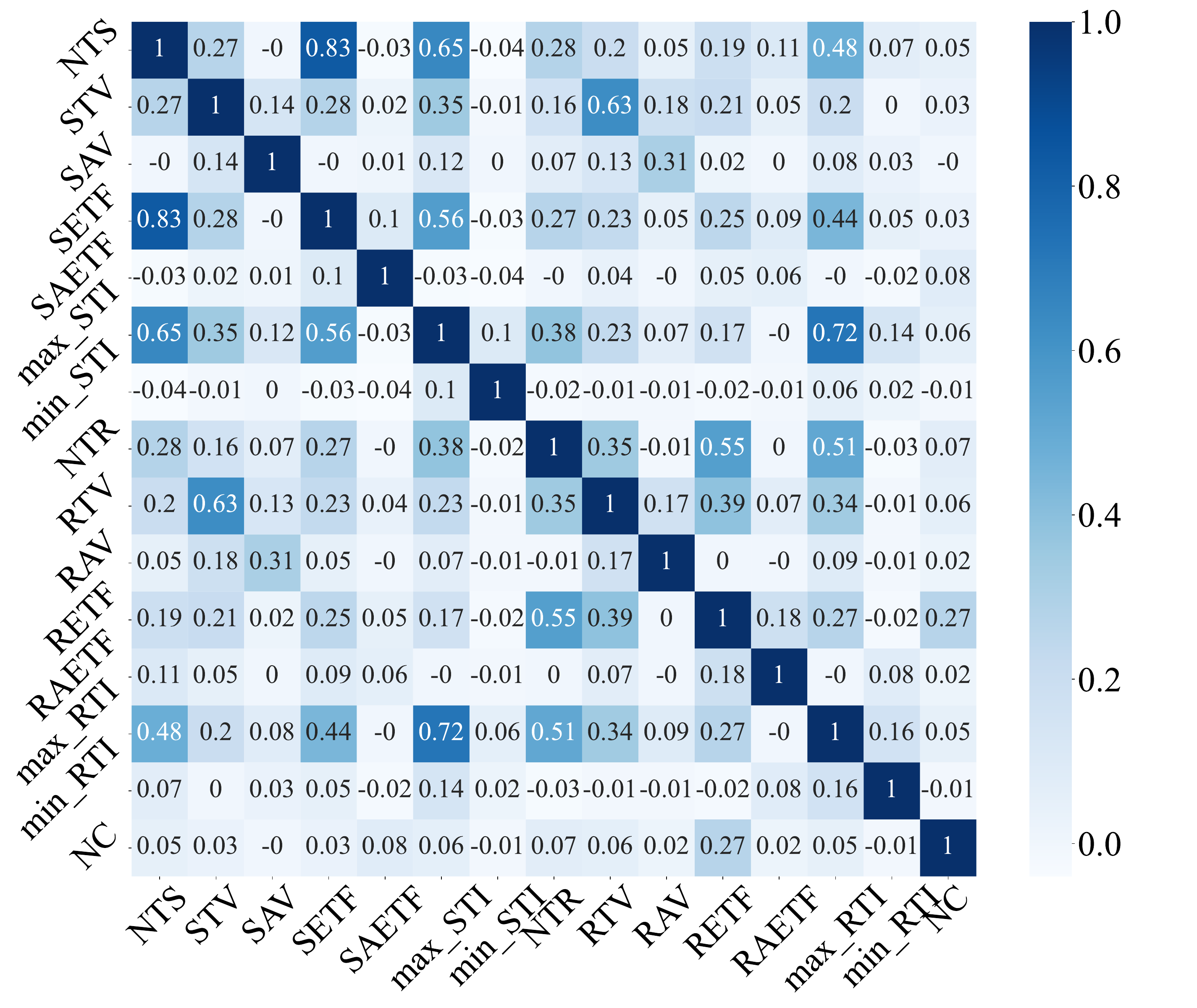}
  \caption{Heat map of 15-dimensional attribute correlation of node features.}
  \label{figureheat}
\end{figure}


We set the baseline methods as follows: the graph embedding methods are set to a walk length of 30, a walk count of 200, and an embedding dimension of 64. Average pooling is used to obtain an embedding representation of the graph. GNN-based methods are stacked with two layers of graph networks. The dimension of the GCN hidden layer is set to 64, the dimension of the GAT hidden layer is set to 256, the number of attention headers is set to 8, the dimension of the GIN hidden layer is set to 256, and the hidden layer $\mathrm{I}^{2}$BGNN is set to 128. We choose the average pooling for the pooling layers of GCN, GAT, and GIN, as well as the maximum pooling for the pooling layer of $\mathrm{I}^{2}$BGNN. For the other baseline methods, we select the optimal configuration from corresponding papers.

\subsection{Irreplaceability of Component (RQ1)}
\subsubsection{Node feature importance validation} 
\begin{figure}[htbp]
  \centering
  \includegraphics[width=8.5cm]{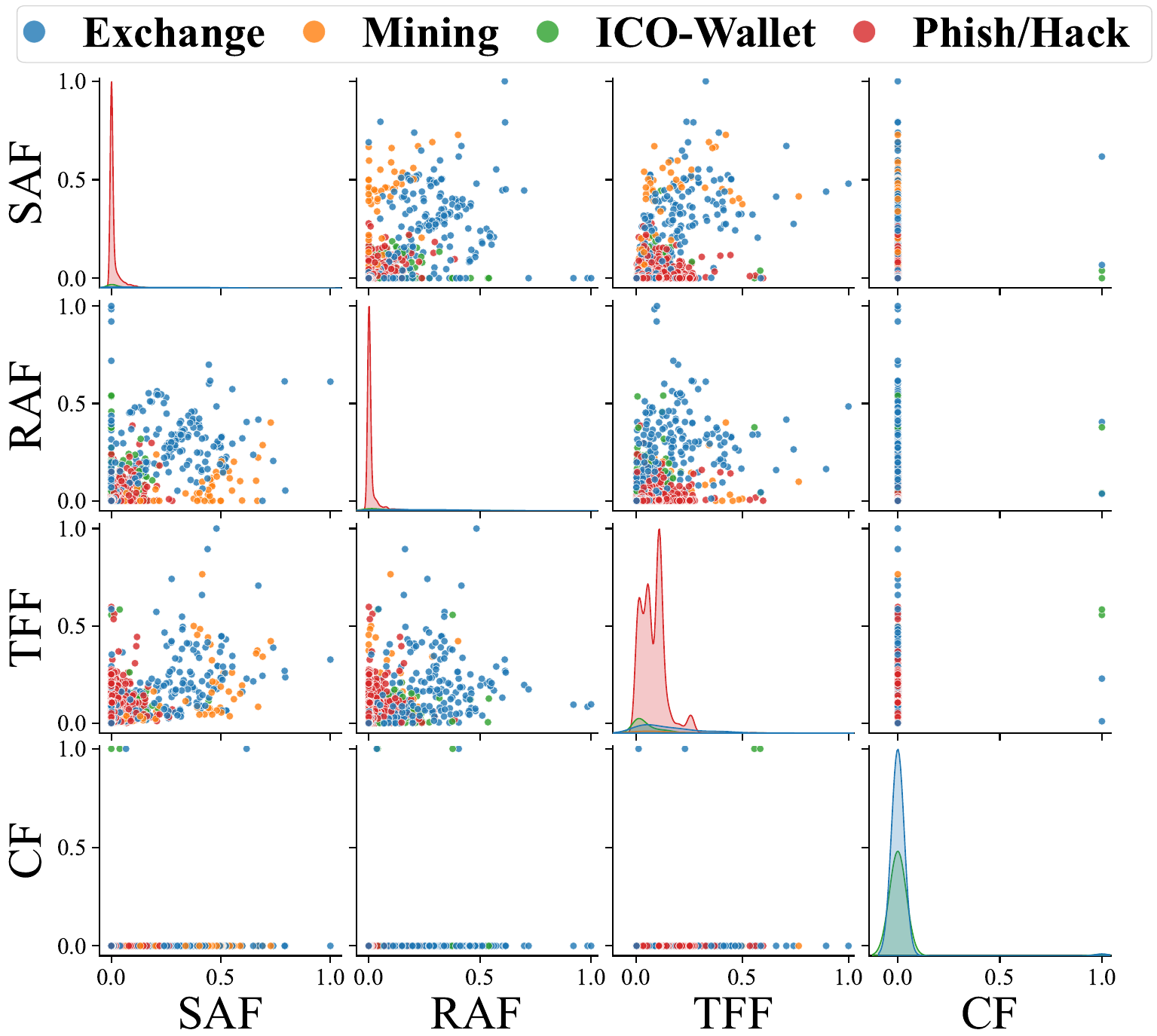}
  \caption{Scatter plot of account category features.}
  \label{figure3}
\end{figure}
We generate 15-dimensional node features for each subgraph node. To verify whether node features are valid, we separate the node features according to their categories and explore whether there is a correlation between the features pairwise. Figure \ref{figureheat} presents a heat map of feature correlation coefficients, reflecting the degree of association between different features. Meanwhile, 
we normalize each of the 15-dimensional features individually. Then, features within the same category are normalized again to obtain the four account category features: sender account feature (SAF), receiver account feature (RAF), transaction fee feature (TFF), and contract feature (CF). The scatter plot distribution of account category features is shown in Figure \ref{figure3}, with the histograms of each category displayed along the diagonal. The above figures show that various distribution patterns are expressed among account types. There is no redundant feature with a strong correlation between the 15-dimensional account features and the four account category features we constructed, which can be directly used for subsequent training.

\begin{figure*}[htbp]
  \centering
  \includegraphics[width=\linewidth]{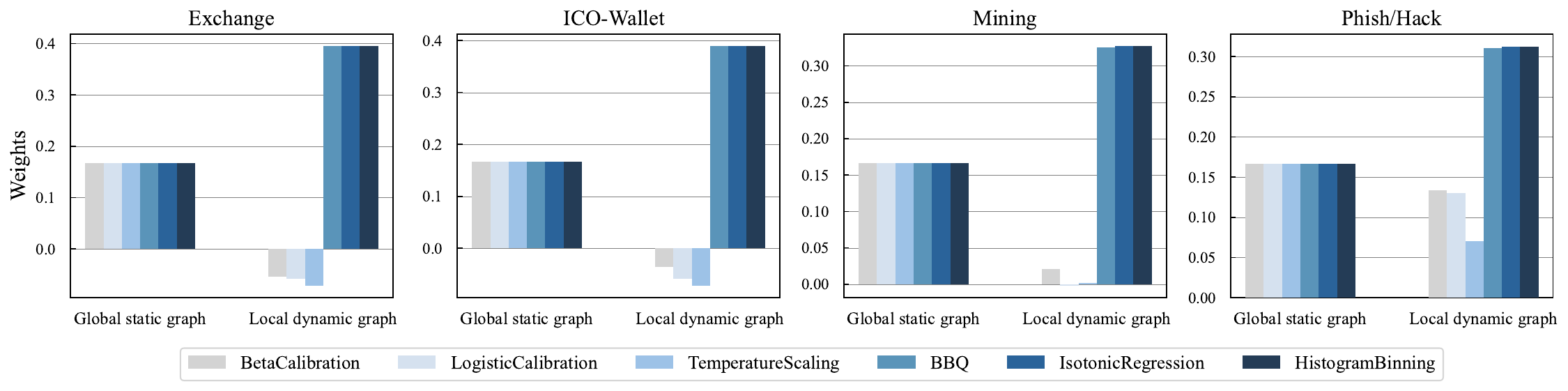}
  \caption{Different types of accounts use different calibration methods to obtain weight proportion data adaptively.}
  \label{figure4}
\end{figure*}

\begin{figure*}[htbp]
  \centering
  \includegraphics[width=\linewidth]{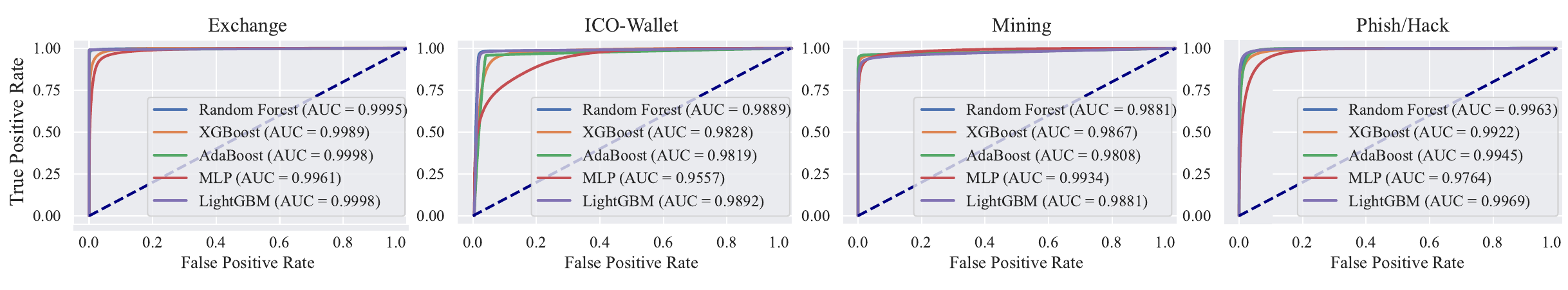}
  \caption{ROC curves for different classifiers selected on four different accounts.}
  \label{figure5}
\end{figure*}
\subsubsection{Evaluation of six confidence calibration methods}
In Section ~\ref{sec:section3.3}, we use six different confidence calibration methods and evaluate the calibration performance of the methods with ECE reduction. To understand the influence of each calibration method in our proposed adaptive calibration approach, we visualize the normalized weights of ECE reductions for GSG and LDG across four different account types, as shown in Figure \ref{figure4}. We conclude that all six calibration methods calibrate to the same extent for the GSG but perform very differently for the LDG. The first three columns of each group of bars in the figure are parametric calibration methods, and the last three groups are non-parametric calibration methods. The statistics show that the proportion of the non-parametric calibration method is larger than that of the parametric calibration method, that is, the non-parametric calibration method has a stronger correction. In exchange, ico-wallet, and mining, there are cases where parametric calibration methods receive negative weights, i.e., the ECE of the corresponding method is increased compared to the ECE without that calibration method. We analyze the reason for this because parametric calibration methods are usually more likely to overfit the calibration data than non-parametric calibration methods, especially when sample sizes are relatively small. In addition, we believe that the similar performance of the confidence calibration methods on the GSG may be due to the ECE metric's weakness in capturing the variance of the predicted values. Additional calibration assessment metrics could be investigated for calculation in subsequent work.

\subsubsection{Classifier selection and performance evaluation}
\label{sec:section4.2}
DBG4ETH adopts LightGBM as the classifier for weighted prediction probabilities after confidence calibration, which integrates the strong learning abilities of both models to identify Ethereum accounts under the best combination. Specifically, after confidence calibration, the weighted calibrated probabilities of GSG and LDG are obtained, and these two types of probabilities are used as input data for LightGBM. To better select appropriate classifiers, we also used MLP, random forest~\cite{29}, AdaBoost~\cite{30}, and XGBoost~\cite{31} for model training in our experiments. The four graphs in Figure \ref{figure5} represent the ROC curves of the five classifiers for different account categories. The purple line in the figure shows the ROC curve for LightGBM, and we can see that the calibrated classification results of LightGBM outperform the other four classifiers across all four account categories.

\begin{table*}[htbp]
  \fontsize{7.8}{11.5}\selectfont 
  \setlength{\tabcolsep}{0pt} 
  	\rmfamily  
    \setlength{\abovecaptionskip}{0pt}%
    \setlength{\belowcaptionskip}{0pt}%
  \caption{Performance comparison of experimental results among DBG4ETH and baselines.}
  \label{tab:4}
  \begin{tabular}{|c|c|cccc|cccc|cccc|cccc|}
    \hline
     & \multicolumn{1}{c|}{\multirow{3}{*}{\textbf{Methods}}} & \multicolumn{16}{c|}{\textbf{Datasets}}\\
     \cline{3-18}
     &  &\multicolumn{4}{c|}{\textbf{Exchange}} & \multicolumn{4}{c|}{\textbf{ICO-Wallet}} & \multicolumn{4}{c|}{\textbf{Mining}} & \multicolumn{4}{c|}{\textbf{Phish/Hack}}\\
    \cline{3-18}
     & & \textbf{Precision} & \textbf{Recall} & \textbf{$F_1$} & \textbf{Accuracy} & \textbf{Precision} & \textbf{Recall} & \textbf{$F_1$} & \textbf{Accuracy} & \textbf{Precision} & \textbf{Recall} & \textbf{$F_1$} & \textbf{Accuracy} & \textbf{Precision} & \textbf{Recall} & \textbf{$F_1$} & \textbf{Accuracy} \\
    \hline
    1 & DeepWalk~\cite{1} & 73.75 & 81.94 & 77.63 & 75.54 & 67.81 & 82.68 & 74.51 & 72.04 & 70.59 & 80.00 & 75.00 & 76.47 & 71.91 & 52.89 & 60.95 & 88.77 \\
    2 & Node2Vec~\cite{3} & 74.55 & 81.30 & 77.78 & 74.10 & 68.29 & 58.33 & 62.92 & 64.52 & 58.83 & 76.92 & 66.67 & 70.59 & 82.68 & 41.77 & 55.50 & 88.36 \\
    \hline
    3 & GCN(w/o node feature) & 40.56 & 46.10 & 43.15 & 36.8 & 53.83 & 50.96 & 52.36 & 39.67 & 36.87 & 42.11 & 39.32 & 35.69 & 40.97 & 50.00 & 45.04 & 45.03 \\
    4 & GCN~\cite{6} & 81.33 & 79.22 & 80.26 & 78.86 & 69.93 & 68.27 & 69.09 & 67.60 & 87.78 & 86.84 & 87.31 & 86.76 & 71.74 & 55.22 & 62.41 & 55.29 \\
    5 & GAT(w/o node feature) & 50.00 & 50.00 & 50.00 & 34.55 & 35.69 & 44.75 & 39.71 & 49.04 & 33.33 & 25.00 & 28.57 & 32.73 & 40.97 & 50.00 & 45.04 & 45.03 \\
    6 & GAT~\cite{7} & 84.61 & 83.12 & 83.86 & 82.93 & 70.73 & 69.23 & 69.97 & 68.66 & 78.27 & 76.32 & 77.28 & 75.90 & 88.35 & 76.23 & 81.84 & 80.44 \\
    7 & GIN (w/o node feature) & 25.00 & 50.00 & 33.33 & 33.33 & 55.25 & 50.96 & 53.02 & 38.38 & 47.37 & 32.15 & 38.30 & 24.32 & 50.00 & 45.03 & 47.39 & 40.97 \\
    8 & GIN~\cite{8} & 82.77 & 81.17 & 81.96 & 80.94 & 25.00 & 50.00 & 33.33 & 33.33 & 83.93 & 76.32 & 79.94 & 74.91 & 86.13 & 81.11 & 83.54 & 83.29 \\
    9 & GraphSAGE\cite{45} & 94.23 & 93.62 & 93.53 & 93.53 & 87.25 & 87.10 & 87.08 & 87.08 & 82.58 & 82.58 & 82.58 & 82.58 & 89.07 & 80.12 & 83.63 & 83.63 \\
    10 & APPNP\cite{36} & 82.09 & 80.76 & 80.46 & 80.46 & 85.52 & 85.48 & 85.48 & 85.48 & 69.70 & 69.70 & 69.57 & 69.57 & 59.88 & 51.08 & 48.00 & 48.00 \\
    11 & GRIT\cite{43} & 73.33 & 53.85 & 48.94 & 48.94 & 50.86 & 50.21 & 51.61 & 51.61 & 23.91 & 50.00 & 47.83 & 47.83 & 55.89 & 56.41 & 73.36 & 73.36 \\
    \hline
     12 & Trans2Vec~\cite{5} & 69.23 & 84.38 & 76.06 & 75.54 & 80.96 & 64.15 & 71.58 & 70.97 & 80.00 & 84.21 & 82.05 & 79.41 & 79.47 & 48.44 & 60.19 & 88.77 \\
    13 & $\mathrm{I}^{2}\mathrm{BGNN}$(w/o node feature) & 81.84 & 81.82 & 81.82 & 81.82 & 80.71 & 80.52 & 80.49 & 80.49 & 78.95 & 78.95 & 78.95 & 78.95 & 88.55 & 79.75 & 83.20 & 83.20 \\
    14 & $\mathrm{I}^{2}\mathrm{BGNN}$~\cite{9} & 82.52 & 82.47 & 82.47 & 82.47 & 79.19 & 77.88 & 77.88 & 77.88 & 72.62 & 71.05 & 70.54 & 70.54 & 85.31 & 81.83 & 83.41 & 83.41 \\
    15 & TSGN\cite{44} & 71.87 & 76.16 & 76.04 & 77.14 & 70.73 & 67.00 & 66.73 & 67.00 & 81.25 & 73.05 & 72.34 & 72.50 & 73.75 & 71.30 & 74.77 & 87.67 \\
    16 & Ethident~\cite{10} & 87.14 & 87.55 & 87.23 & 87.23 & 73.33 & 75.00 & 70.97 & 70.97 & 75.00 & 75.00 & 66.67 & 66.67 & 80.28 & 87.80 & 88.93 & 88.93 \\
    17 & TEGDetector~\cite{15} & 85.74 & 85.63 & 85.67 & 85.71 & 80.63 & 81.32 & 80.77 & 81.00 & 86.67 & 84.46 & 84.65 & 85.00 & 81.33 & 80.42 & 80.86 & 88.63 \\
    {18} &  BERT4ETH ~\cite{17}& {77.25} & {76.89} & {76.69} & {76.74} & {78.44} & {78.59} & {77.53} & {77.53} & {85.00} & {81.67} & {82.37} & {83.33} & {81.65} & {86.40} & {83.59} & {88.08} \\
    \hline
    {19} &  DBG4ETH & \textbf{99.46} & \textbf{99.03} & \textbf{99.51} & \textbf{99.46} & \textbf{97.58} & \textbf{96.80} & \textbf{97.19} & \textbf{97.18} & \textbf{97.73} & \textbf{95.24} & \textbf{97.56} & \textbf{97.72} & \textbf{97.43} & \textbf{98.98} & \textbf{98.42} & \textbf{97.43} \\
    \hline
    &Improve. & \textcolor{red}{(5.23)} & \textcolor{red}{(5.41)} & \textcolor{red}{(5.98)} & \textcolor{red}{(5.93)} & \textcolor{red}{(10.33)} & \textcolor{red}{(9.70)} & \textcolor{red}{(10.11)} & \textcolor{red}{(10.10)} & \textcolor{red}{(9.95)} & \textcolor{red}{(8.40)} & \textcolor{red}{(12.91)} & \textcolor{red}{(10.96)} & \textcolor{red}{(8.36)} & \textcolor{red}{(11.18)} & \textcolor{red}{(9.49)} & \textcolor{red}{(8.50)} \\
    \hline
  \end{tabular}
\end{table*}

\subsection{Effectiveness of DBG4ETH (RQ2)} 
To validate the performance of our proposed DBG4ETH method, we select two node embedding-based methods, six graph-based methods and six Ethereum account identification methods on the same dataset to compare the experimental results, as shown in Table \ref{tab:4}. For the GNN-based baselines, we conducted two experiments for each method: one without the generated node features and another with adding 15-dimensional node features. The statistics show that performance improves after adding node features compared to without them. For example, GCN's $F_1$-score can be increased by 16.73\%, reaching 47.99\%. Further observation of the statistic in Table \ref{tab:4} reveals that our proposed DBG4ETH obtains state-of-the-art experimental results, which are 5.23\% to 12.91\% higher than the rest of the baseline methods, proving that our well-designed DBG4ETH method can capture account behavior patterns more accurately.

\subsection{Ablation Study (RQ3)}
To verify the contribution of each module in DBG4ETH to the final classification results, we conduct three sets of ablation experiments and record the $F_1$-score in Table \ref{tab:5}. 
\subsubsection{Verify the ability of account classification using GSG or LDG alone}
Firstly, we verify the ability of account classification using GSG or LDG alone. We perform the following ablations: DBG4ETH \textit{without GSG} and \textit{without LDG}. Using either graph alone does not surpass the experimental results obtained from combining both graphs. In terms of $F_1$-score, our method improves by up to 40.52\% compared to using GSG alone and by up to 32.67\% compared to using LDG alone. Moreover, our method achieves a good result of 97.19\% on the $F_1$-score, even for the ico-wallet, which performs poorly for the two benchmark models.

\subsubsection{Verify the adaptive calibration module}
To validate the adaptive calibration module, we design two sets of experiments: (1) DBG4ETH \textit{without calibration}: We remove all calibration methods, three parametric calibration, and three non-parametric calibration methods (denoted as w/o calibration, w/o Param. calibration and w/o Non-param. calibration) through three sets of experiments respectively, which are used to verify whether adding calibration methods helps to further infer account identity. (2) DBG4ETH \textit{without adaptive calibration}: We remove adaptive parametric calibration, adaptive non-parametric calibration, and adaptive calibration (denoted as w/o Ada. Param. calibration, w/o Ada. Non-param. calibration and w/o Ada. calibration) through three sets of experiments to validate the necessity of adding adaptive weights for some calibration methods individually and for all calibration methods. The experimental results in Table \ref{tab:5} show that incorporating calibration methods and assigning weights to different calibration methods significantly improves the results.
\subsubsection{Verify the classifier performance}

\begin{table}[htbp]
  \fontsize{8}{11.5}\selectfont 
  \setlength{\tabcolsep}{0.5pt} 
  \caption{Performance comparison of model ablation experiments.}
  \label{tab:5}
  \begin{tabular}{|c|c|c|c|c|}
    \hline
    \textbf{Models} & \textbf{Exchange} & \textbf{ICO-Wallet} & \textbf{Mining} & \textbf{Phish/Hack} \\
    \hline
    w/o GSG & 87.50 & 56.67 & 80.00 & 90.83 \\
    w/o LDG & 78.72 & 64.52 & 75.00 & 93.44 \\
    \hline
    w/o calibration & 94.23 & 83.05 & 78.05 & 97.11 \\
    w/o Param. calibration & 99.03 & 89.76 & 68.00 & 98.31 \\
    w/o Non-param. calibration & 97.58 & 98.21 & 93.02 & 98.24 \\
    \hline
    w/o Ada. Param. calibration & 99.50 & 88.89 & 97.56 & 98.30 \\
    w/o Ada. Non-param. calibration & 97.08 & \textbf{98.28} & 75.00 & 98.41 \\
    w/o Ada. calibration & 98.49 & 98.26 & 97.54 & 98.23 \\
    \hline
    w/o LightGBM & 96.13 & 91.80 & 81.63 & 98.29 \\
\hline
    DBG4ETH & \textbf{99.51} & 97.19 & \textbf{97.56} & \textbf{98.42} \\
    \hline
  \end{tabular}
\end{table}
The validation of the classifier performance is presented in Section ~\ref{sec:section4.2}. The results of DBG4ETH \textit{without LightGBM} in Table \ref{tab:5} are classified with MLP instead of LightGBM classifier.

As a result, it is observed that our proposed DBG4ETH performs best in classifying Ethereum accounts, and it correctly classifies account types even for those with poor initial results or small datasets.

\subsection{Impact Of Dynamic Cryptocurrency Market On DBG4ETH (RQ4)}
Considering that the cryptocurrency market is dynamically changing, we add two new account identities (bridge and defi) to verify the practical strength of DBG4ETH. 

\begin{figure}[htbp]
\begin{subfigure}{0.25\textwidth}
    \centering
    \includegraphics[width=\textwidth]{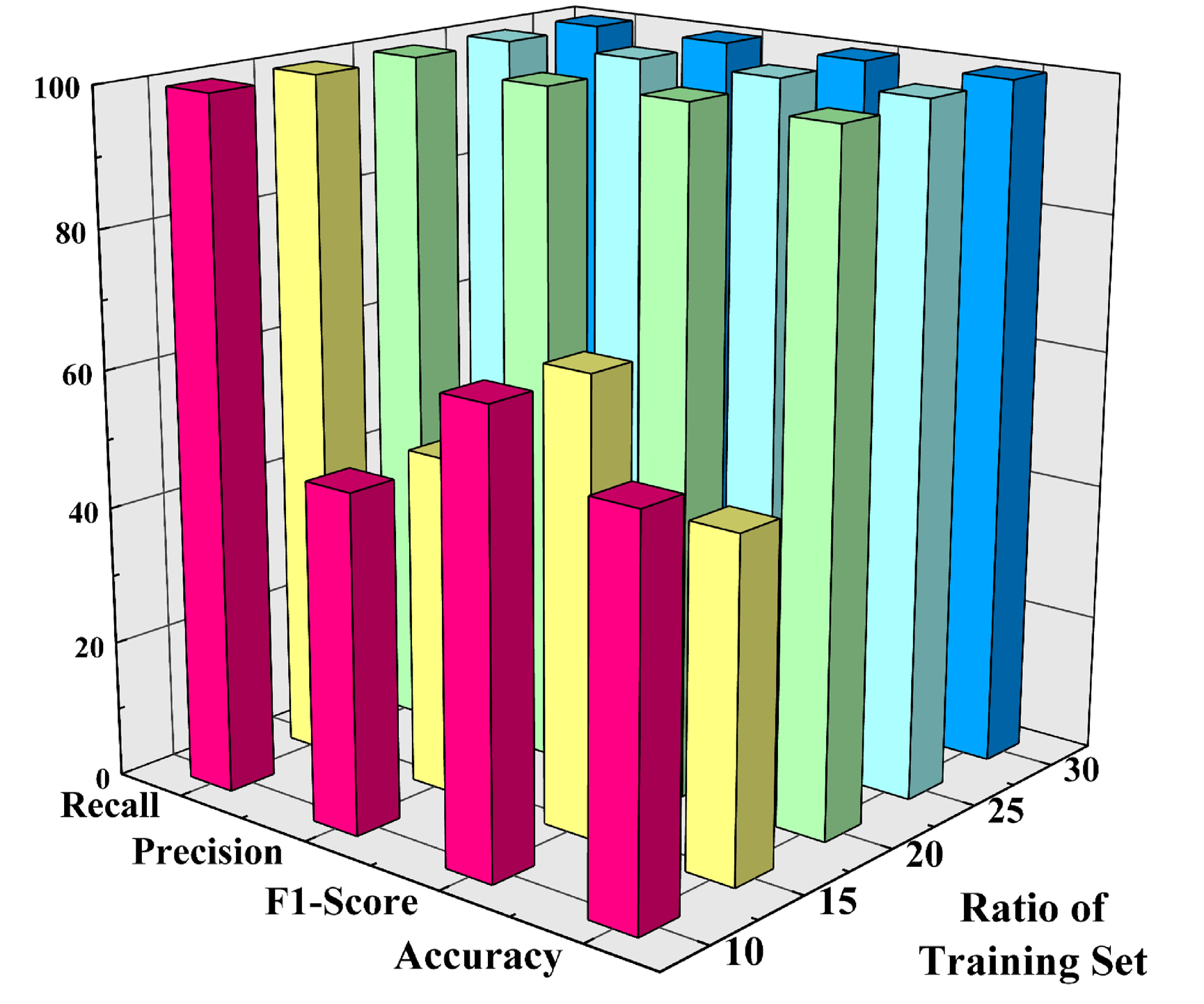}
    \caption{Bridge.}
    \label{fig:6a}
  \end{subfigure}%
  \begin{subfigure}{0.25\textwidth}
    \centering
    \includegraphics[width=\textwidth]{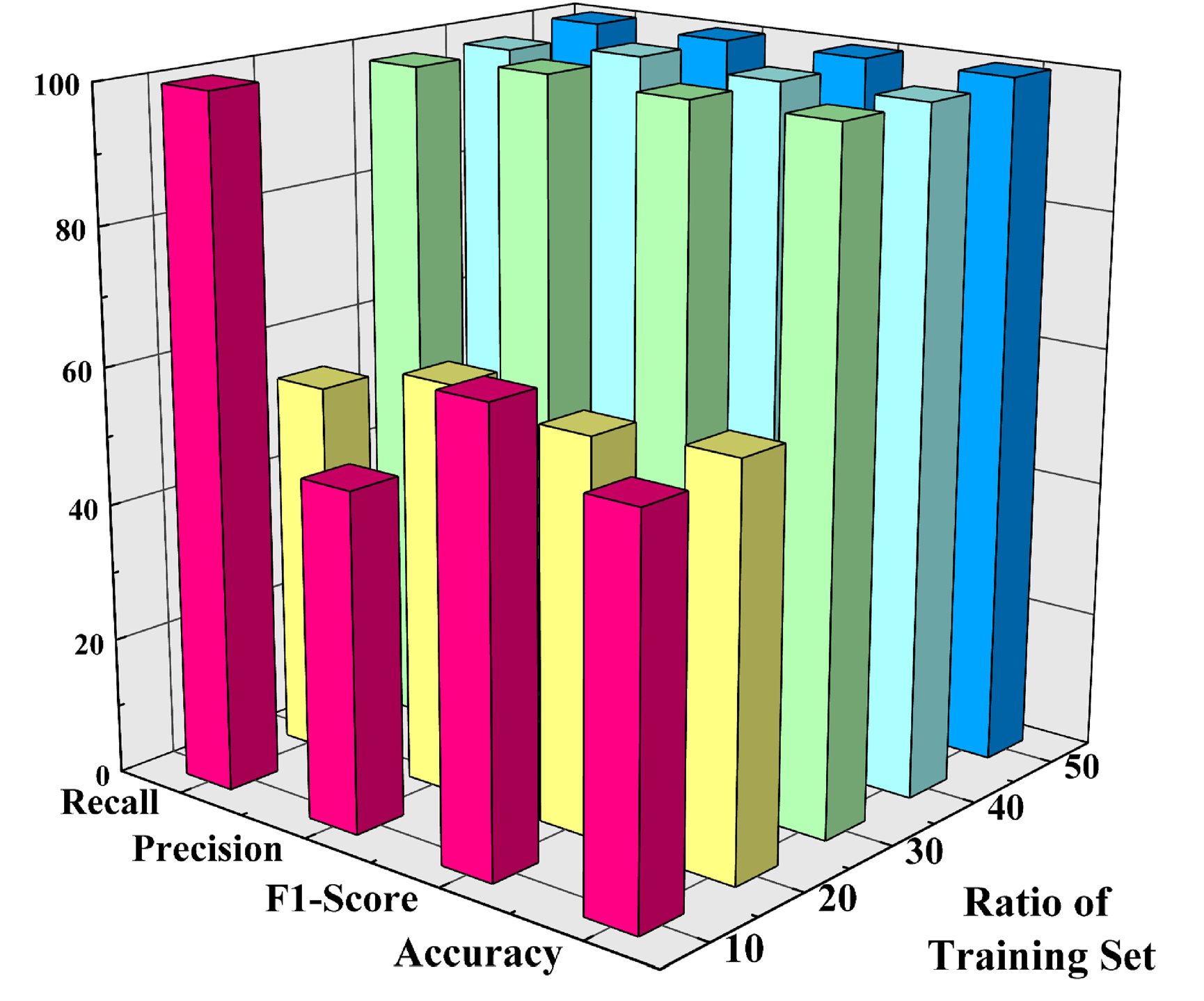}
    \caption{DeFi.}
    \label{fig:6b}
  \end{subfigure}%
  \caption{The impact of training set size on model performance.}
\label{figure6}
\end{figure}

\begin{table}[htbp]
\fontsize{8}{11.5}\selectfont 
  \setlength{\tabcolsep}{9.5pt} 
 \caption{Account classification results on bridge.}
  \label{tab:6}
\begin{tabular}{|c|cccc|}
\hline
\multirow{2}{*}{Models}  & \multicolumn{4}{c|}{Bridge} \\ 
\cline{2-5} 
& \multicolumn{1}{c|}{Pricision} & \multicolumn{1}{c|}{Recall} & \multicolumn{1}{c|}{$F_1$} & \multicolumn{1}{c|}{Accuracy}  \\ 
\hline
\multicolumn{1}{|c|}{DeepWalk~\cite{1}}  & \multicolumn{1}{c|}{65.62} & \multicolumn{1}{c|}{63.64}       & \multicolumn{1}{c|}{64.62}   & \multicolumn{1}{c|}{63.49}   \\
\hline
\multicolumn{1}{|c|}{GCN~\cite{6}} & \multicolumn{1}{c|}{93.75}          & \multicolumn{1}{c|}{92.86}       & \multicolumn{1}{c|}{93.30}   & \multicolumn{1}{c|}{92.82}   \\
\multicolumn{1}{|c|}{GIN~\cite{8}} & \multicolumn{1}{c|}{91.67}          & \multicolumn{1}{c|}{90.00}       & \multicolumn{1}{c|}{90.83}   & \multicolumn{1}{c|}{89.90}   \\
\multicolumn{1}{|c|}{GraphSAGE~\cite{45}}  & \multicolumn{1}{c|}{96.05}          & \multicolumn{1}{c|}{95.71}       & \multicolumn{1}{c|}{95.88}   & \multicolumn{1}{c|}{95.71}   \\
\hline
\multicolumn{1}{|c|}{$\mathrm{I}^{2}\mathrm{BGNN}$~\cite{9}} & \multicolumn{1}{c|}{97.30}          & \multicolumn{1}{c|}{97.14}       & \multicolumn{1}{c|}{97.14}   & \multicolumn{1}{c|}{97.14}   \\
\multicolumn{1}{|c|}{Ethident~\cite{10}} & \multicolumn{1}{c|}{94.59}          & \multicolumn{1}{c|}{96.54}       & \multicolumn{1}{c|}{97.22}   & \multicolumn{1}{c|}{97.14}   \\ 
\multicolumn{1}{|c|}{TEGDetector~\cite{15}} & \multicolumn{1}{c|}{76.39}          & \multicolumn{1}{c|}{75.79}       & \multicolumn{1}{c|}{76.67}   & \multicolumn{1}{c|}{76.67}\\
\multicolumn{1}{|c|}{BERT4ETH~\cite{17}} & \multicolumn{1}{c|}{97.62}          & \multicolumn{1}{c|}{97.06}       & \multicolumn{1}{c|}{97.27}   & \multicolumn{1}{c|}{97.30} \\ 
\hline
\multicolumn{1}{|c|}{DBG4ETH} & \multicolumn{1}{c|}{\textbf{98.64}}          & \multicolumn{1}{c|}{\textbf{100}} & \multicolumn{1}{c|}{\textbf{99.32}}   & \multicolumn{1}{c|}{\textbf{99.32}}   \\
\hline
\end{tabular}
\end{table}

\begin{table}[htbp]
\fontsize{8}{11.5}\selectfont 
  \setlength{\tabcolsep}{9.5pt} 
 \caption{Account classification results on defi.}
  \label{tab:7}
\begin{tabular}{|c|cccc|}
\hline
\multirow{2}{*}{Models}  & \multicolumn{4}{c|}{DeFi} \\ 
\cline{2-5} 
& \multicolumn{1}{c|}{Pricision} & \multicolumn{1}{c|}{Recall} & \multicolumn{1}{c|}{$F_1$} & \multicolumn{1}{c|}{Accuracy}  \\ 
\hline
\multicolumn{1}{|c|}{DeepWalk~\cite{1}}  & \multicolumn{1}{c|}{63.33} & \multicolumn{1}{c|}{59.38}       & \multicolumn{1}{c|}{61.29}   & \multicolumn{1}{c|}{61.90}   \\
\hline
\multicolumn{1}{|c|}{GCN~\cite{6}} & \multicolumn{1}{c|}{93.75}          & \multicolumn{1}{c|}{92.86}       & \multicolumn{1}{c|}{93.30}   & \multicolumn{1}{c|}{92.82}   \\
\multicolumn{1}{|c|}{GIN~\cite{8}} & \multicolumn{1}{c|}{96.05}          & \multicolumn{1}{c|}{95.71}       & \multicolumn{1}{c|}{95.88}   & \multicolumn{1}{c|}{95.71}   \\
\multicolumn{1}{|c|}{GraphSAGE~\cite{45}}  & \multicolumn{1}{c|}{96.05}          & \multicolumn{1}{c|}{95.71}       & \multicolumn{1}{c|}{95.88}   & \multicolumn{1}{c|}{95.71}   \\
\hline
\multicolumn{1}{|c|}{$\mathrm{I}^{2}\mathrm{BGNN}$~\cite{9}} & \multicolumn{1}{c|}{97.30}          & \multicolumn{1}{c|}{97.14}       & \multicolumn{1}{c|}{97.14}   & \multicolumn{1}{c|}{97.14}   \\
\multicolumn{1}{|c|}{Ethident~\cite{10}} & \multicolumn{1}{c|}{94.59}          & \multicolumn{1}{c|}{96.54}       & \multicolumn{1}{c|}{97.22}   & \multicolumn{1}{c|}{97.14} \\
\multicolumn{1}{|c|}{TEGDetector~\cite{15}} & \multicolumn{1}{c|}{63.84}          & \multicolumn{1}{c|}{64.03}       & \multicolumn{1}{c|}{63.33}   & \multicolumn{1}{c|}{63.33}\\
\multicolumn{1}{|c|}{BERT4ETH~\cite{17}} & \multicolumn{1}{c|}{97.22}          & \multicolumn{1}{c|}{96.15}       & \multicolumn{1}{c|}{96.57}   & \multicolumn{1}{c|}{96.66} \\ 
\hline
\multicolumn{1}{|c|}{DBG4ETH} & \multicolumn{1}{c|}{\textbf{100}}    & \multicolumn{1}{c|}{\textbf{98.63}}      & \multicolumn{1}{c|}{\textbf{99.31}}   & \multicolumn{1}{c|}{\textbf{99.32}}   \\
\hline
\end{tabular}
\end{table}

Table \ref{tab:6} and \ref{tab:7} present the experimental results for the account types bridge and defi. By comparing our method with baseline approaches, we demonstrate that DBG4ETH achieves near-perfect prediction accuracy even when dealing with novel account types. To further investigate whether the excellent performance is due to the limited size of the labeled dataset, we conducted additional experiments by varying the ratio of training samples from 10\% to 50\% of the total dataset. Interestingly, our model performs exceptionally well with only a relatively small subset of the training data. Specifically, with just 20\% of the training set for bridge and 30\% for defi, our method can achieve optimal predictive performance. The specific experimental results are shown in Figure \ref{figure6}. This is attributed to our model's ability to retain global information from historical transactions and analyze the evolution of local transactions, thereby enabling deeper insights into user-related activities.

\subsection{Hyperparameters Sensitivity Analysis (RQ5)}
In this subsection, we mainly conduct sensitivity analysis on the key parameters in DBG4ETH. These parameters primarily include the four hyperparameters that influence the generation of graph views in GSG (i.e., $P_{e,1}$, ${P_{e,2}}$, ${P_{f,1}}$, ${P_{f,2}}$) and the number of pooling layers in LDG.


\subsubsection{Sensitivity Analysis of Hyperparameters in GSG Encoding} In GSG, $P_{e,1}$, ${P_{e,2}}$ represent the probabilities of removing edges in the two generated graph views, while ${P_{f,1}}$, ${P_{f,2}}$ control the degree of masking node attributes in the two graph views. These four hyperparameters introduce noise to unimportant edges and node features, thereby controlling the degree of augmentation in topology and node attributes. For simplicity in visualization, we set ${P_{e,1}=P_{e,2}}$ and ${P_{f,1}=P_{f,2}}$, with the adjustment range between 0 and 1. In the sensitivity analysis, we vary only these four parameters, while keeping other parameters the same as in Section \ref{sec:section5.1}. The results of the ico-wallet dataset are shown in Figure \ref{figure9} (a). From the figure, we can observe that when the values of the two types of parameters are relatively small \( (< 0.4) \), our model is not sensitive to changes in these hyperparameters, demonstrating the robustness of DBG4ETH to hyperparameter perturbations. However, when the parameters are set too large, the original graph is severely disrupted, leading to too many isolated nodes in the generated graph, negatively affecting the learning of node representations.


\subsubsection{Sensitivity Analysis of Hyperparameters in LDG Encoding} 
\begin{figure}[htbp]
\begin{subfigure}{0.25\textwidth}
    \centering
    \includegraphics[width=\textwidth]{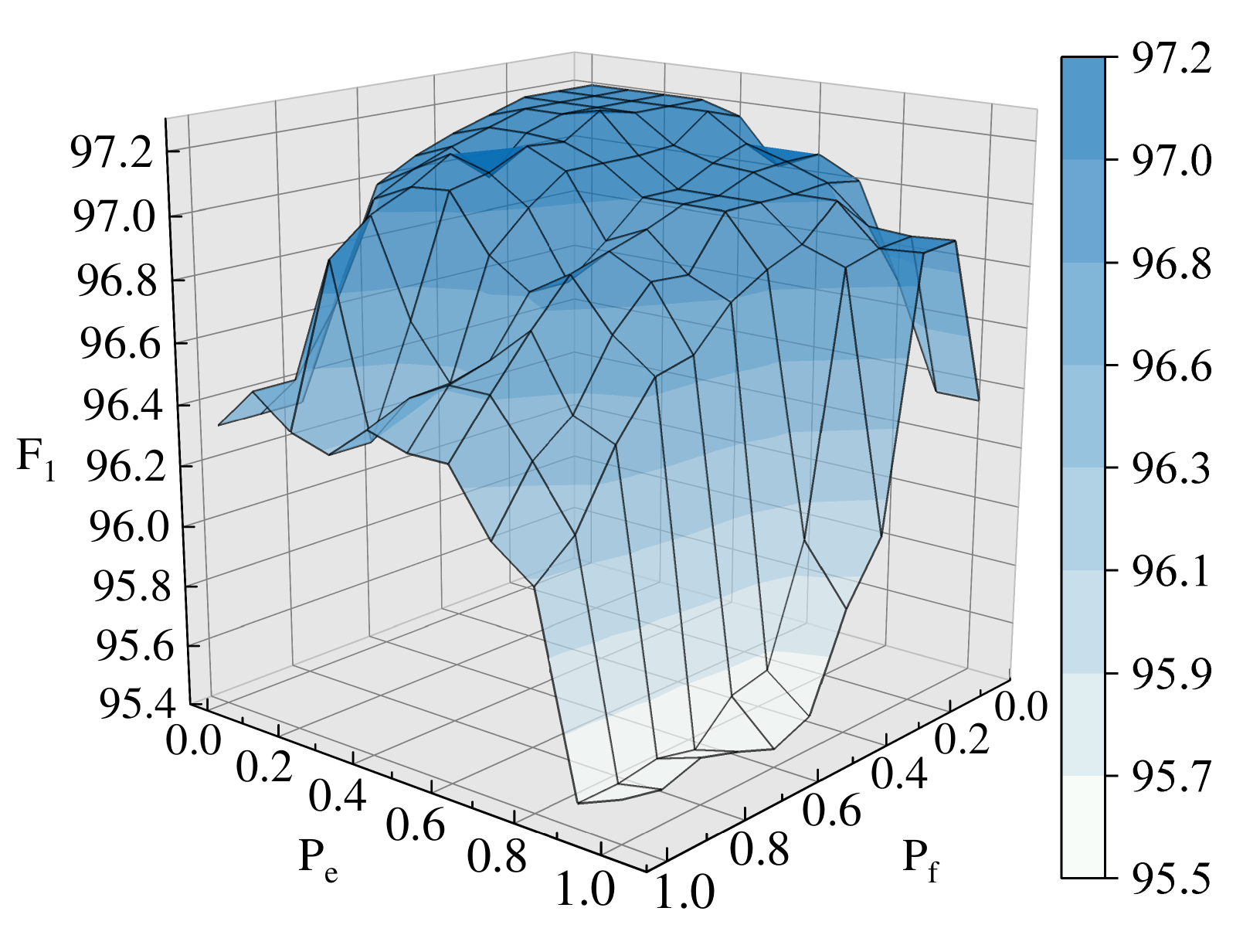}
    \caption{GSG encoder.}
    \label{fig:6a}
  \end{subfigure}%
  \begin{subfigure}{0.25\textwidth}
    \centering
    \includegraphics[width=\textwidth]{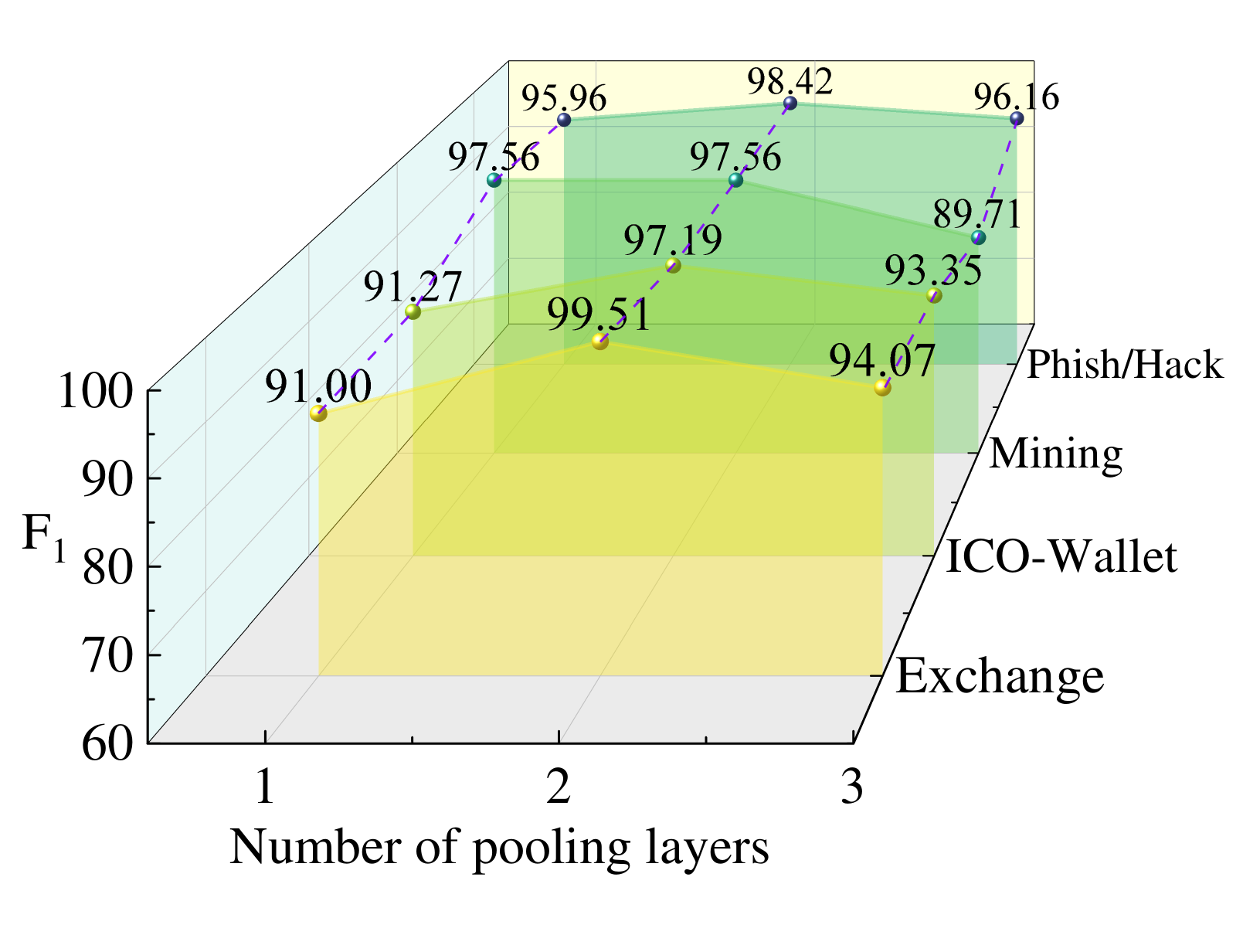}
    \caption{LDG encoder.}
    \label{fig:6b}
  \end{subfigure}%
  \caption{Effect of hyperparameters in GSG and LDG encoder on the experimental results.}
\label{figure9}
\end{figure}

In LDG encoding, the most critical parameter is the number of pooling layers. Each time slice contains maximum nodes $N=2000$, and the pooling rate is set to 0.1. Therefore, the maximum number of pooling layers that can be applied is 3. We conducted experiments on four different account datasets, adjusting the number of pooling layers from 1 to 3. The experimental results are shown in Figure \ref{figure9} (b). From the figure, we can observe that our model achieves the best performance when the number of pooling layers is set to 2, and the overall change in the number of layers has a relatively small impact on the experimental results.

\section{CONCLUSION}
The financial crimes in DeFi have caused enormous economic losses to innocent customers and hindered the development of Web 3.0. 
The major challenge in blockchain regulation is to analyze the real identity of customers, including the account de-anonymization inference under the scarcity of account labels. 
In this paper, we propose a novel Ethereum account identification framework, dubbed DBG4ETH, which combines the advantages of the global static graph and local dynamic graph to perceive different accounts' behavior patterns and achieve a SOTA de-anonymization inference effect. Meanwhile, we add an adaptive confidence calibration module to improve trustworthiness in real application scenarios. Experimental results show that our method achieves better results than baselines. In addition, future work should focus on account de-anonymization tasks under privacy-protecting services, such as Tornado Cash, which obscure transaction analysis by disrupting fund flow tracking, making it a crucial and impactful area of research.

\section{Acknowledgements}
This work is partially supported by the National Key Research and Development Program of China under Grant 2021YFB2700300, the National Natural Science Foundation of China (62141605, 62372493), the Beijing Natural Science Foundation (Z230001), the Beijing Advanced Innovation Center for Future Blockchain and Privacy Computing (GJJ-23-001, GJJ-23-002), and the science and technology project of State Grid Corporation of China (5400-202255416A-2-0-ZN), and the National Natural Science Foundation of China under Grant, and National Key R\&D Program of China under Grant 2022ZD0116800, and the China Postdoctoral Fellowship Fund 373500, and the Beihang Dare to Take Action Plan KG16336101.

\bibliography{references.bib} 
\bibliographystyle{IEEEtran}

\end{document}